\documentclass[a4paper]{article}
\usepackage{comment}
\usepackage{subfigure}
\usepackage{amssymb,amsthm,amscd, amsbsy, array}
\usepackage{amsmath,bm,cite,color}
\usepackage{tikz}
\usepackage{graphicx}
\newcommand{\mathsym}[1]{{}}

\usepackage{epsfig}
\usepackage{epsf}
\usepackage{rotating}
\usepackage{color}
\newfont{\tenmsb}{msbm10 scaled\magstep1}

\let\ssection=\section\renewcommand{\section}
{\setcounter{equation}{0}\ssection}

\newcommand{\mb}{\mathbf}

\newcommand{\lf}{\left (}

\newcommand{\rg}{\right )}

\def\smallover#1/#2{\hbox{$\textstyle{#1\over#2}$}}
\def\beq{\begin{equation}}
\def\eeq{\end{equation}}
\def\beq{\begin{equation}}
\def\eeq{\end{equation}}
\def\bea{\begin{eqnarray}}
\def\eea{\end{eqnarray}}
\newcommand{\nn}{\nonumber}
\def\lf{\left(}
\def\rg{\right)}
\def\lq{\left[}
\def\rq{\right]}
\def\lgr{\left\{}

\def\p{{\partial}}
\def\p{{\partial}}

\def\vD{{\vec D}}
\def\vE{{\vec E}}

\def\vr{{\vec r}}

\def\vcE{{\vec \cE}}

\def\bj{{\bf j}}
\def\bk{{\bf k}}

\def\bn{{\bf n}}

\def\bphi{\boldsymbol{\phi}}

\def\cA{{\mathcal A}}

\def\cE{{\mathcal E}}

\def\*{{\star}}

\def\hr{{\hat r}}

\def\hx{{\hat x}}
\def\hy{{\hat y}}
\def\hz{{\hat z}}

\def\freccia{{ \rightarrow }}

\def\xe{{{\'e }}}

\def\aa{{{\`a }}}

\begin{document}
\title{Light Scattering by Cholesteric Skyrmions}

\author{G. De Matteis \dag \footnote{e-mail: giovanni.dematteis@istruzione.it}, $\quad$  D. Delle Side \dag \footnote{e-mail:nico.delleside@unisalento.it}, $\quad$  L. Martina \dag \ddag \footnote{e-mail:martina@le.infn.it}, $\quad$ V. Turco \dag \footnote{ e-mail: vito.turco@live.com}
\\$*$ IISS   "V. Lilla", MIUR,  Francavilla Fontana (BR) Italy \\ \dag Dipartimento di Matematica e Fisica, Universit\aa del Salento\\  \ddag  INFN, Sezione di  Lecce  \\Via per Arnesano, C.P. 193 I-73100 Lecce, Italy\\ }

\maketitle

\begin{abstract}
  We study the light scattering by localized quasi planar excitations of a Cholesteric Liquid Crystal known as spherulites. Due to the anisotropic optical properties of the medium and the peculiar shape of the excitations, we   quantitatively  evaluate the cross section of the axis-rotation of polarized light.  Because of the complexity of the system under consideration, first we give a simplified, but analytical, description of the spherulite and we compare the Born approximation results in this setting with those obtained by resorting to  a numerical  exact  solution. The effects  of changing values of the driving external static electric (or magnetic) field is considered. Possible applications of the phenomenon are envisaged.
   \end{abstract}
 
  \section{Introduction}\label{sec:intro}

    In the last few years great efforts  have been done in developing new materials  for  opto-electronics and photonics applications. A relevant role in this work has been  played by  the liquid crystals (LC)  physics  \cite{Luckhurst2017,Chigrinov2010} for a quite long time. In fact,   nowadays LC are  widely used in all types of display applications and  their unique nonlinear electro-optical  properties  make them suitable material for non-display applications, like optical filters and switches, beam-steering devices, spatial light modulators, optical wave-guiding, lasers \cite{Coles} and optical nonlinear components \cite{Beeckman}. On the other hand,  a wide interest was deserved by a  variety of new  2-dimensional structures like  \textit{cholesteric fingers} \cite{OSWALD200067}, and 3-dimensional ones, like  nematicons  \cite{Assanto:2016aa}, \textit{cholesteric bubbles} or \textit{spherulites} \cite{doi:10.1080/02678299208029010,doi:10.1080/02678299108035502,POL:POL180180709}, the latter appear in quasi-2D layers of Chiral Liquid Crystals (CLCs) with homeotropic anchoring on the confining surfaces. Those  textures  have been  studied from a theoretical point of view \cite{Leonov,noiProc} and we would like to consider them for their potential opto-technological applications. Thus the aim of the present paper is  to evaluate the possibility to exploit spherulites, isolated or in lattice arrangements  \cite{key-2,carboni}, as electric/magnetically driven switches for light beams, propagating in the liquid crystal. 

Spherulites in CLCs  share  some properties with the 2D skyrmions in magnetic systems \cite{Romming636,PhysRevLett.87.037203}. In fact, these isolated axisymmetric states are stabilized by specific interactions imposed by the underlying molecular handedness,  however they  are more sensible to the external fields and may possess slow modulations in a preferred direction.   Thus, a continuum  model can be derived in the framework of the Frank-Oseen theory \cite{deGennes,Stewart}, from which one can write the respective equilibrium equations. By applying external fields and  imposing anchoring boundary conditions \cite{1,2},  the free  helicoidal equilibrium  can be deformed into   new structures such as skyrmions \cite{9,12}, which are stabilized by topological  conservation laws. The theory also describes the cholesteric fingers \cite{3,8}, or helicoids, with defect disclination type, which can be  described, at least in some approximate setting,  in terms of integrable nonlinear equations \cite{key-2,17},  stabilized both by  topological and non-topological conservation laws.  Carboni et al.  \cite{carboni} detected a phase transition between the two textures, strongly depending on the thickness of the confining cell. They showed that the texture changes are  driven  by temperature through a parameter $\zeta$ proportional to the thickness and to a proper chirality parameter. Samples of different thickness displayed the textural changes at different temperatures but for the same value of $\zeta$. 
However here we limit ourselves to the sferulites/skyrmion case. 

 The paper is organized as follows. In Sec. \ref{sec:skyrme} we introduce the continuum elastic model of the CLC, we obtain the corresponding equilibrium equations and analyse the skyrmion (spherulite) solutions, either by analytical or numerical methods. In Sec. \ref{sec:diffusion} we introduce the problem of light diffusion by  a spherulite. In Sec. \ref{sec:Born} we provide perturbative solutions for the light scattering equations derived in Sec. \ref{sec:diffusion}. In particular in \ref{sec:Bornout} we compute the cross section of the conversion process of incoming polarized light in the incidence plane into the  outgoing polarized light in the perpendicular direction. Analogously,   in \ref{sec:Bornin} we consider the complementary problem of the change of polarization axis from the direction orthogonal to the liquid crystal  to the in plane direction. Finally, in the Conclusions we summarise our results and address some possible experimental realizations.

\section{Skyrmions in chiral liquid crystals}\label{sec:skyrme}
A  LC 
 is described by a uni-modular director field $\mathbf{n}\lf \mathbf{r} \rg$ belonging to $\mathbb{RP}^2$\cite{deGennes,Stewart}, which   in polar representation is
\begin{equation}
\mathbf{n}(\mb{r})=(\sin\theta(\mathbf{r})\cos\psi(\mathbf{r}), \sin\theta(\mathbf{r})\sin\psi(\mathbf{r}), \cos\theta(\mathbf{r})), \qquad - \mathbf{n} \equiv \mathbf{n}. \label{directorpolar}
\end{equation}
In the bulk  a CLC  director field $\mathbf{n}\lf \mathbf{r} \rg$ is governed by the Frank-Oseen  free energy density 
\bea
\omega_{FO}=    \frac{K_1}{2}(\nabla\cdot\mathbf{n})^2+\frac{K_2}{2}(\mathbf{n}\cdot\nabla\times\mathbf{n}-q_0)^2 +\frac{K_3}{2} (\mathbf{n}\times\nabla\times\mathbf{n})^2\nonumber\\
+\frac{(K_2+K_4)}{2}\nabla\cdot [(\mathbf{n}\cdot\nabla)\mathbf{n}-\nabla\cdot\mathbf{n}-(\nabla\cdot\mathbf{n})\mathbf{n}] -\frac{\varepsilon}{2}(\mathbf{n}\cdot\mathbf{E})^2 
, \label{fomega}
\eea
where  $q_0$ is the chirality parameter of the cholesteric phase, the positive reals $K_1$, $K_2$, $K_3$, $ K_4$ are the Frank elastic constants, which we set to be $ K = K_1 = K_2 = K_3, \quad K_4=0 $ for sake of simplicity. The last term  in (\ref{fomega}) represents the interaction energy density associated with a spatially uniform external  static electric field $\mathbf{E}$, or equivalently a magnetic field  $\mathbf{H}$, along the $\mb{k}$ direction.  Of course, in the presence of the external electric (magnetic)  field, the general rotational symmetry is broken and reduced to rotations around the direction of $\mathbf{E}$ ($\mathbf{H}$).   In the absence of anchoring conditions, the  field $\mathbf{n}\lf \mathbf{r} \rg$ would form a cholesteric helix with axis orthogonal to $\mathbf{E}$  ($\mathbf{H}$).  However, supposing  the CLC confined within the region $\mathcal{B}=\lbrace (x,y,z)\in\mathbb{R}^3,  \mid z\mid\leq \dfrac{L}{2}\rbrace$,  the translational symmetry in the direction of $\mb{k}$ is broken and the interaction of the CLC with the planar bounding surfaces  can be encoded  by the Rapini and Papoular\cite{rapini}  additional surface energy contribution 
\beq
\omega_s=\frac{1}{2} K_s (1+\alpha(\mb{n}\cdot\bm{\nu})^2),
\eeq
where $K_s,\hspace{.1cm}\alpha>0$ and $\bm{\nu}$ being the unit outward normal to the boundary surface.  Strong homeotropic anchoring is obtained for $K_s\to\infty$, which corresponds to the  Dirichlet boundary conditions 
$ \mathbf{n}\lf x, y, z \pm \frac{L}{2}\rg = \mathbf{k} \equiv  -  \mathbf{k}.\label{surfcond2}$ So  helices are deformed and confined within $\mathcal{B}$ and  possibly extended structures called  helicoids (or  helicons and, sometimes,  \emph{fingers}) or spherulites (also \emph{skyrmions}) may  form, depending on the existence of a preferred direction of perturbations of $\mb{n}$.

In order to find equilibrium configurations of the CLC  we have to minimise the Frank free energy under the appropriate boundary conditions.  We also limit ourselves to axisymmetric isolated solutions. Thus, assuming $\theta=\theta(\rho, z)$ and $\psi=\psi(\phi)$, where $\rho$, $ z$ and $\phi$ are the usual cylindrical coordinates around the axis $\mb{k}$, the solution of minimal energy is given by 
\beq
\psi(\phi)=\phi+\frac{\pi}{2},\hspace{.5cm} \phi\in[0,2\pi]\label{eqpsi2}.
\eeq
and all the admissible equilibrium configurations are  solutions of the dimensionless Boundary Value Problem (BVP)
 \begin{eqnarray}
\label{theta2Dscal}
\frac{\p^2 \theta}{\p z^2}+\frac{\p^2 \theta}{\p \rho^2}+\frac{1}{\rho}\frac{\p \theta}{\p \rho} -\frac{1}{\rho^2}\sin\theta\cos\theta
\mp \frac{4\pi}{\rho}\sin^2\theta-\pi^4\left(\frac{E}{E_0}\right)^2\sin\theta\cos\theta=0,
\vspace{1cm}\\
\begin{cases}
\label{bccases}
&\theta(0,z)= \pi,\vspace{.5cm}\hspace{.5cm}\theta(\infty,z)= 0,\\
&\p_z\theta\left(\rho,\pm\frac{\nu}{2}\right)=\mp 2\pi k_s \sin\theta\left(\rho,\pm \frac{\nu}{2}\right)\cos\theta\left(\rho,\pm \frac{\nu}{2}\right),
\end{cases}
 \end{eqnarray}
where the lengths are  rescaled with respect to the so-called pitch length  $p=\frac{2\pi}{\mid q_0\mid}$. Here, $E_0=\dfrac{\pi \mid q_0\mid}{2} \sqrt{\dfrac{K}{\varepsilon}}$ is the critical unwinding field for the  cholesteric-nematic transition in non-confined CLCs\cite{stewarta}, $\nu=L/p$ is the normalized thickness of the layer and $k_s=K_s/(K q_0)$ the strength of the interaction liquid/ boundary surfaces. The $\mp$ sign in equation \eqref{theta2Dscal} depends on the sign of $q_0$: in the following we take $q_0<0$, with no loss of generality. Moreover, it is convenient to simplify the notation setting $\rho_1^2=\pi^4\left( \dfrac{E}{E_0}\right)^2$.
 System  (\ref{theta2Dscal}-\ref{bccases})   is a 3D perturbed Sine-Gordon type equation:  chirality  and BCs do not allow to integrate it in analytical form. The main deformation comes from the fifth term in (\ref{theta2Dscal}), associated to the chirality of the system.
Thus, the solutions of the BVP (\ref{theta2Dscal}-\ref{bccases})  can be obtained, at least to our knowledge, only by numerical methods. 

However, to get information about  the  shape of a spherulite, one can evaluate the  asymptotic behaviours of the solutions near  $\rho\leadsto 0$ and  $\rho\leadsto\infty$.
Moreover, let us  consider first the pure cylindrical reduction of \eqref{theta2Dscal}, i.e. $\theta_z=0$, which holds when $\nu$ is sufficiently large and modulations in the $z$ variable are discarded. 

\begin{figure}[!htt]
\centering
\subfigure[Contour plot of solution \eqref{linansatz} for $(E/E_0, \nu,k_s)=(1.02, 1.8, 0)$.]
{\includegraphics[width=5cm, height=4cm]{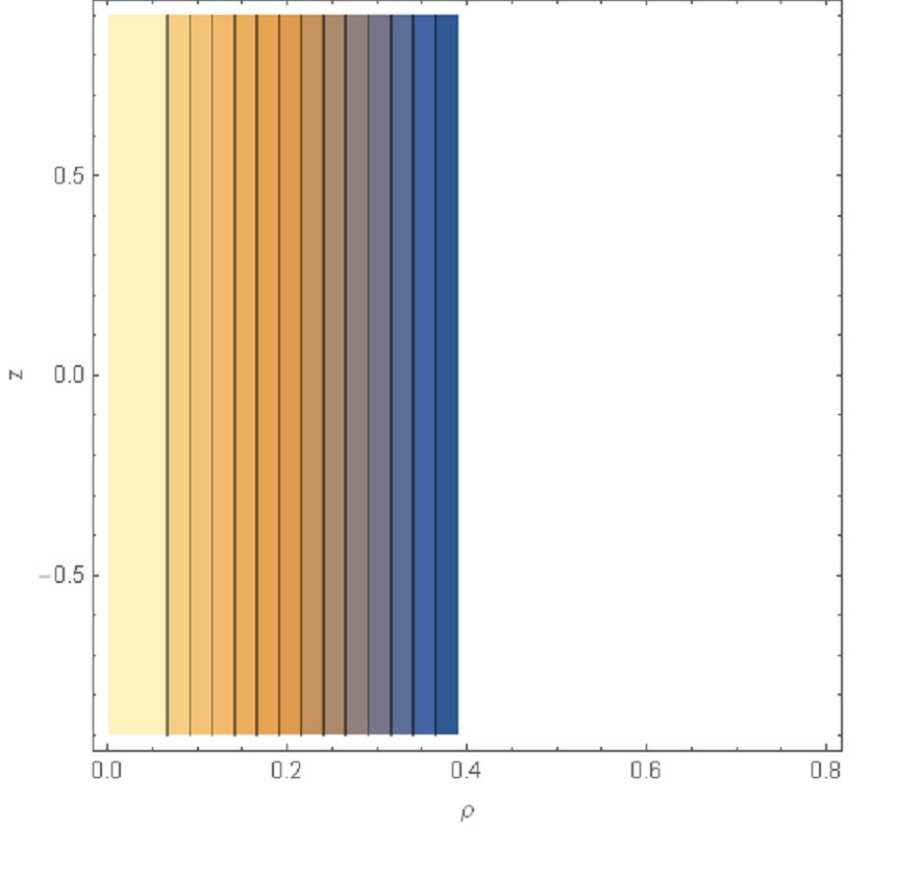}}
\subfigure[Contour plot of solution \eqref{linansatz} for $(E/E_0, \nu,k_s)=(1.02, 1.8, 1.5)$.]
{\includegraphics[width=5cm, height=4cm]{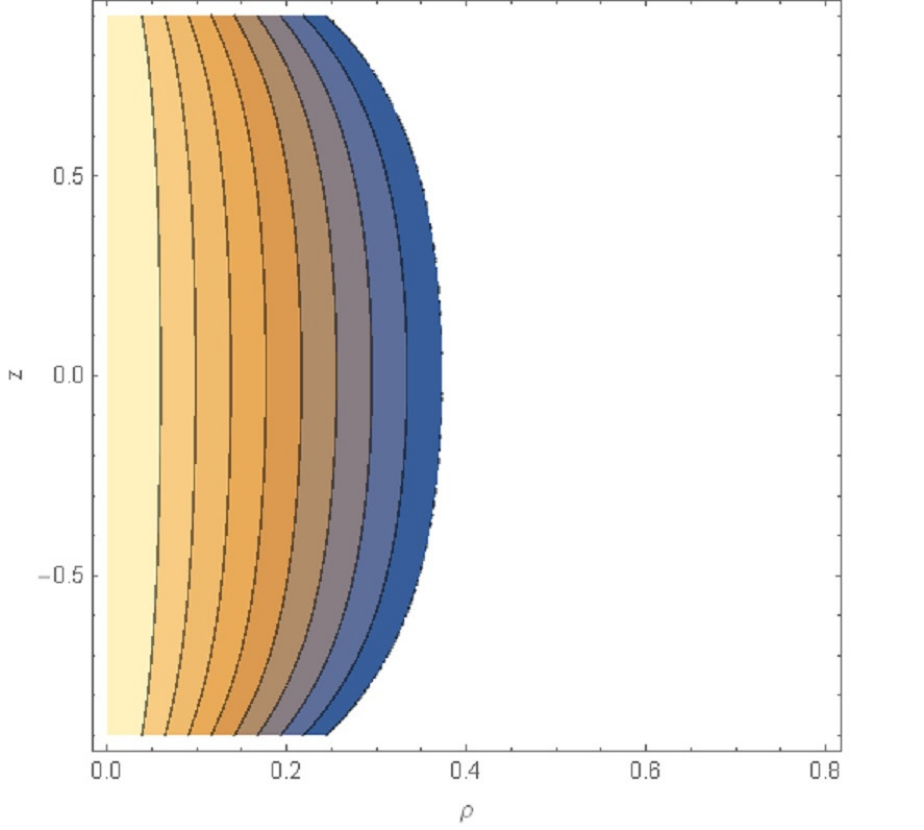}}
\subfigure[Contour plot of solution \eqref{linansatz} for $(E/E_0, \nu,k_s)=(1.02, 1.8, 5)$.]
{\includegraphics[width=5cm, height=4cm]{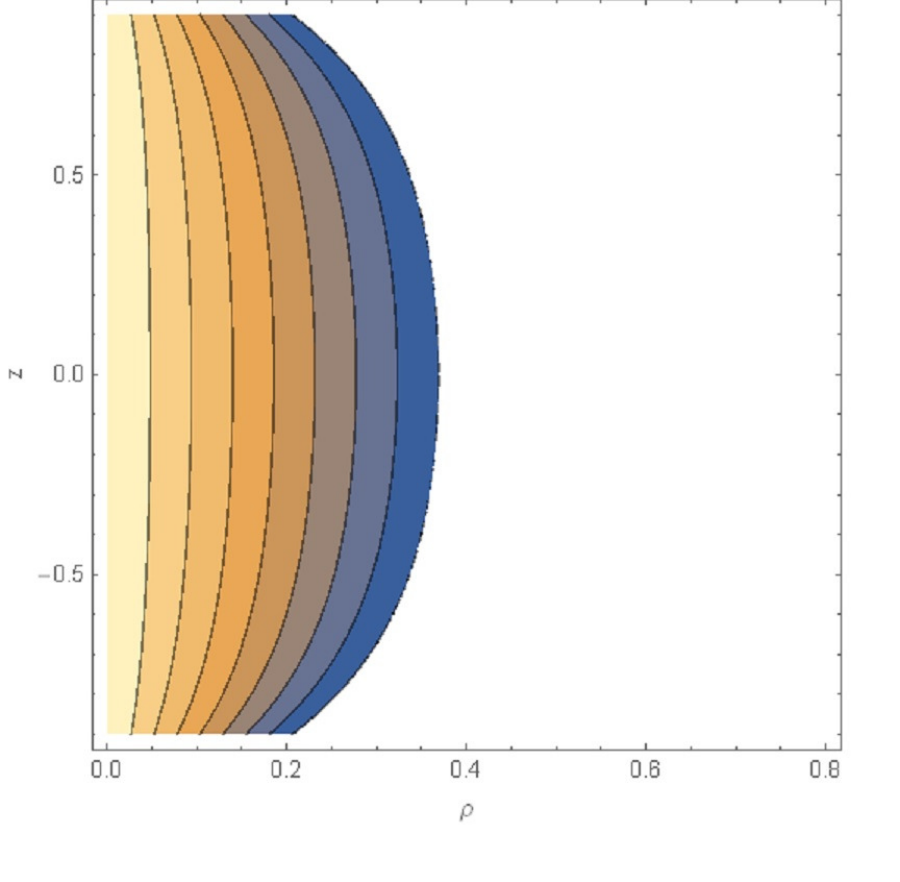}}
\caption{{Contour plot of solution \eqref{linansatz} for three different values of $k_s$ with constant external electric field and cell thickness.}}
\label{fig:approxsfe}
\end{figure}

Near $\rho\leadsto 0$ both the chiral and the electric interaction can be neglected with respect to the other  terms, thus setting   both $q_0\to 0$ and $E\to 0$,  equation \eqref{theta2Dscal} reduces to the  the conformally invariant O(3)-sigma model in polar representation\cite{manton}.
Accordingly, the solutions  near $\rho\leadsto 0$ behave like the  Belavin-Polyakov ones\cite{belp}, namely 
\beq
\label{1BP}
\theta=\arccos\left(\frac{\tilde{\rho}^2-4}{\tilde{\rho}^2+4}\right), \quad \tilde{\rho}=\dfrac{\rho}{\rho_0}
\eeq
 where $\rho_0$
is an arbitrary scale factor due to the conformal invariance.  The fourth and the fifth term in \eqref{theta2Dscal}  break the conformal symmetry.  Thus,
substituting solution \eqref{1BP} in equation \eqref{theta2Dscal} we obtain the   extimation 
\beq
\label{rho0}
\rho_0= \frac{4}{\pi^3} \left(\frac{E_0}{E}\right)^2 = 4\pi \rho_1^2,
\eeq
which can be interpreted as the typical scale of a spherulite. Then, around $\rho= 0$, at the lowest order the solution of (\ref{theta2Dscal}-\ref{bccases}) is approximated by
\beq
\label{ansatzbulk0}
\theta (\rho)= \pi-\frac{\rho}{\rho_0}+O\left(  \left(\frac{\rho}{\rho_0}\right)^3\right).
\eeq
  with $\rho_0$ fixed by \eqref{rho0}. 
 Furthermore, in order to have information also about the modulation in the $z$ direction, we assume as rough approximation of the solution  by deforming \eqref{ansatzbulk0} as 
 \beq
\label{linansatz}
\theta (\rho ,z)= \lgr 
\begin{array}{cc}
  \pi- \frac{\rho}{\rho_0 Z(z)}  &  \rho/Z(z)< \pi\rho_0 \\
0  &   \rho/Z(z)>\pi\rho_0  \end{array}
 \right. , 
\eeq
with $\rho_0$ given by \eqref{rho0}, and replace 
 \eqref{linansatz} into the  Frank-Oseen energy. Its minimisation leads  to   equation
\beq
\label{eleqz}
 Z''(z)- \frac{1}{\pi^2  \rho _1^2} Z(z)+\frac{1}{\pi^2  \rho _1^2}=0,
\eeq
which, upon   imposing the boundary conditions \eqref{bccases},   yields  the  solution 
\beq
\label{zansatz}
Z(z)= 1-\frac{2 \pi k_s  \cosh \left(\frac{ z}{\rho_1
   }\right)}{2 \pi  k_s  \cosh \left(\frac{1}{\rho_1}\frac{  \nu }{2 
   }\right)+\frac{1}{\rho_1} \sinh \left(\frac{1}{\rho_1}\frac{ \nu }{2
   }\right)}.
\eeq
We note that the sizes of the vortices decrease as $\mid z\mid$ and $k_s$ increase, as it can be seen in figure  \ref{fig:approxsfe}.  We will assume hereafter that this is the $z$-modulation of the skyrmion in the entire volume. 

In the asymptotic limit  $\rho\to\infty$  the dominant term  comes from the external electric field, which  affects the shape of Skyrmion by   the reduced  equation  
\beq
\label{sgcil}
\frac{\p^2 \theta}{\p\rho^2}+\frac{1}{\rho}\frac{\p\theta}{\p \rho}-\frac{\rho_1^2}{2} \sin 2\theta=0,
\eeq
which is known as cylindrical Sine-Gordon equation \cite{barone}. The most relevant fact about this equation  is its connection with the celebrated Painlev\xe $\textrm{III}$ equation\cite{ablowitz,McCoy, noiProc}   (see  also \cite{NIST:DLMF} Cap. 32): then it can be  analytically solved. However, in correspondence to the boundary conditions at $\infty$ stated in \eqref{bccases},  this equation has always singular solutions at $\rho\to 0$. Thus the validity of such an approximation is limited to a neighbourhood of $\infty$, where its  asymptotics is 
\beq
\label{asymptlin}
\theta\leadsto c_2\sqrt{\dfrac{\rho_1}{\rho}} \exp\left[-\dfrac{\rho}{\rho_1}\right].
\eeq
This result is sufficiently similar to the one obtained in linear approximation, which leads to first order modified Bessel functions of second kind which have almost analogous asymptotics.

The above results give us useful  indications about the shape of the spherulite/skyrmion, but many important details are missed. In fact, to have a good account of them and to estimate the goodness of the approximations made above, we need to perform numerical calculations on  the BVP described by  (\ref{theta2Dscal}-\ref{bccases}). To this aim,  we  use the standard central finite difference discretisation and the Newton-Raphson method \cite{recipes, leveque}, inizialized by  the  shooting method for the planar reduction of the system  (i.e. $\theta_z=0$).

\begin{figure}
\centering
\includegraphics[width=10cm, height=5cm]{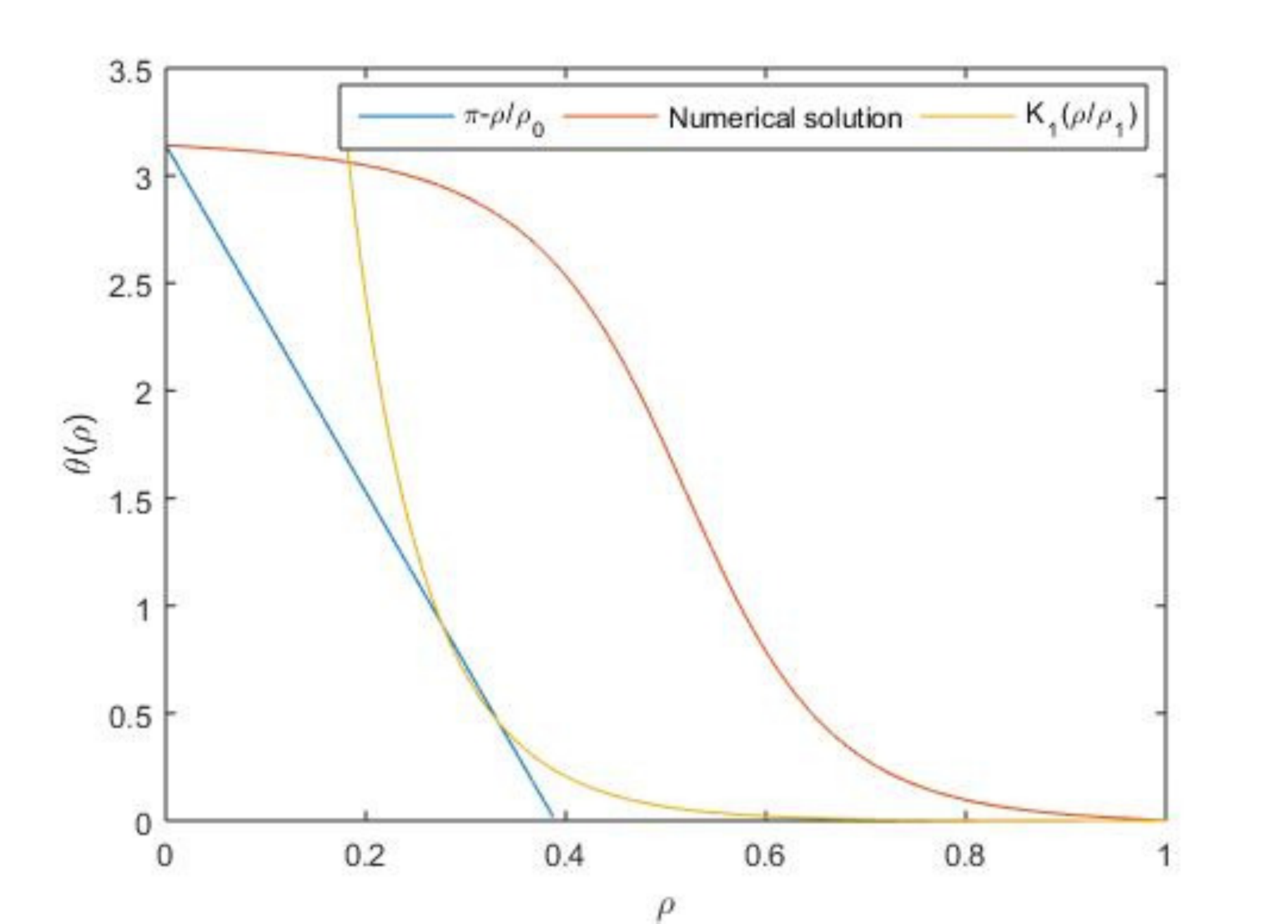}
\label{fig:comparison1.02Ana}
\caption{Comparison between the numerical solution of \eqref{theta2Dscal} and the analytical linear approximations for $\frac{E}{E_0}=1.02$.}
\end{figure}

\begin{figure}
\centering
\includegraphics[width=10cm, height=5cm]{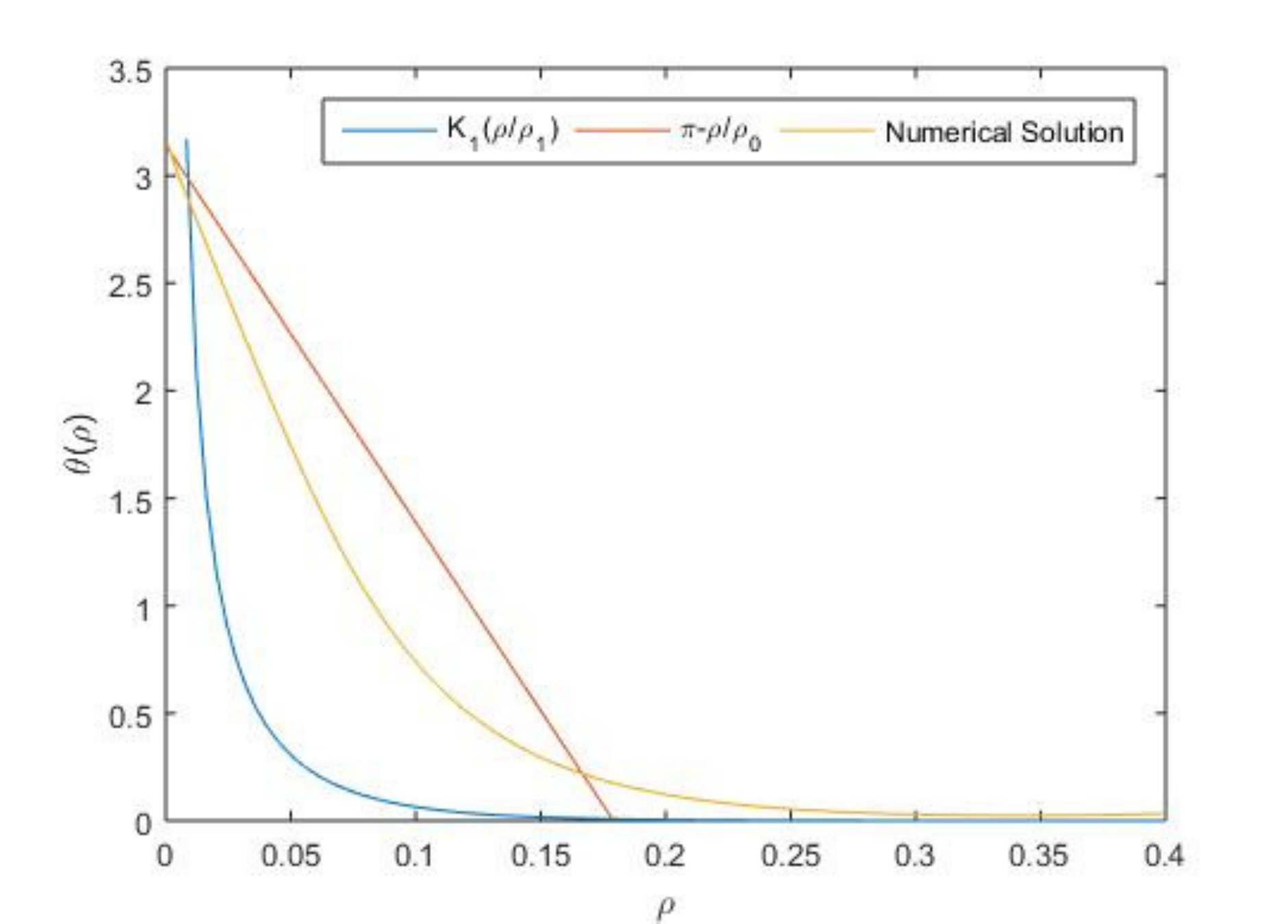}
\label{fig:comparison1.5Ana}
\caption{Comparison between the numerical solution of \eqref{theta2Dscal}  and the analytical linear approximations for $\frac{E}{E_0}=1.5$}
\end{figure}

It turns out  that for sufficiently large  electric fields, i.e. $\frac{E}{E_0} > 1$ the linear approximations  matches  with the numerical solution quite closely, as represented in fig. \ref{fig:comparison1.02Ana}. On the other hand, the approximations become very rough for relatively weak fields , i.e. $\frac{E}{E_0}  \approx 1$, as shown in fig. \ref{fig:comparison1.5Ana}. As far as  the numerical cases considered in the present work, this behaviour denotes the underestimation of the  chiral term in the linear approximation, in particular at the intermediate scales $\rho_1\leq \rho  \leq \rho_0$.  


\begin{figure}[!htt]
\centering
\subfigure[$\left(\dfrac{E}{E_0}\right) =1.02,k_s=0.1$]
{\includegraphics[width=.48\linewidth, height=5cm]{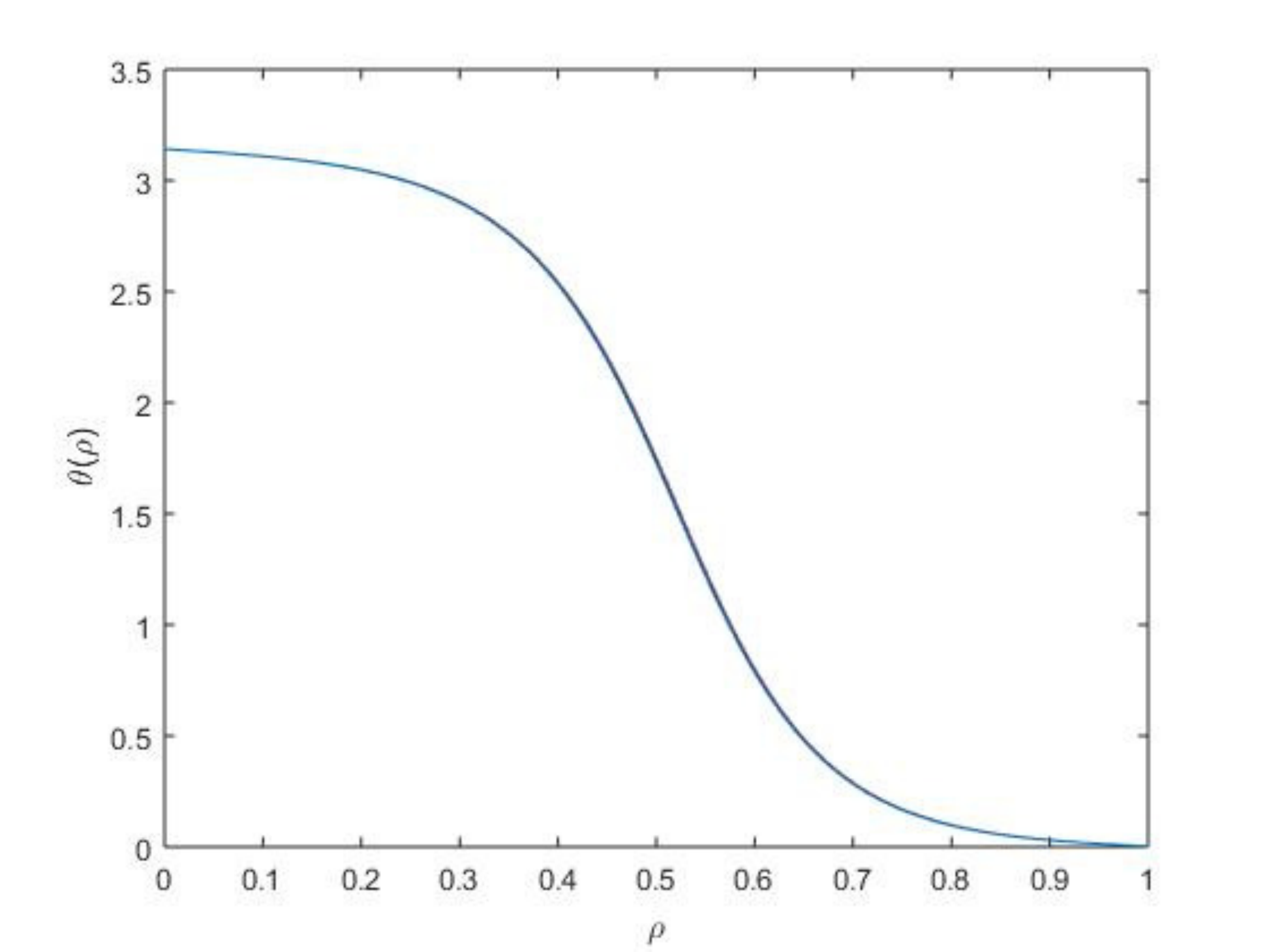}}
\subfigure[$\left(\dfrac{E}{E_0}\right) =1.02, k_s=6$]
{\includegraphics[width=.48\linewidth, height=5cm]{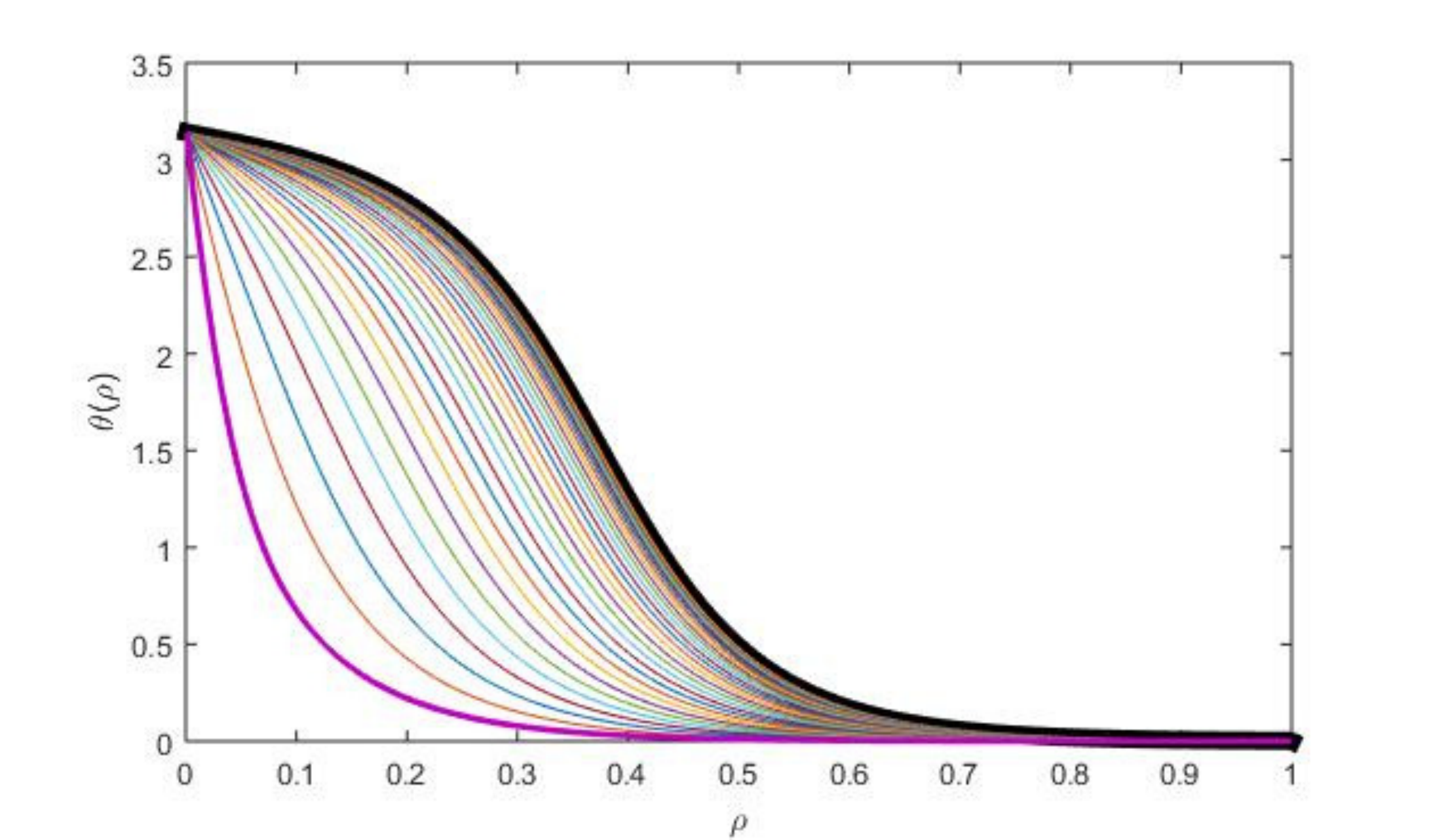}}
\caption{Profiles $\theta (\rho)$ for $E/E_0=1.02$. Different curves refer to different values of $\mid z\mid$.  Bold curves have to be referred  to $z =0$ (the black one) and to $\mid z	\mid=\nu/2$ (the purple one).}
\label{fig:profiles1}
\end{figure}

\begin{figure}[!ht]
\centering
\subfigure[$\left(\dfrac{E}{E_0}\right) =1.5,k_s=0.1$]
{\includegraphics[width=.48\linewidth, height=5cm]{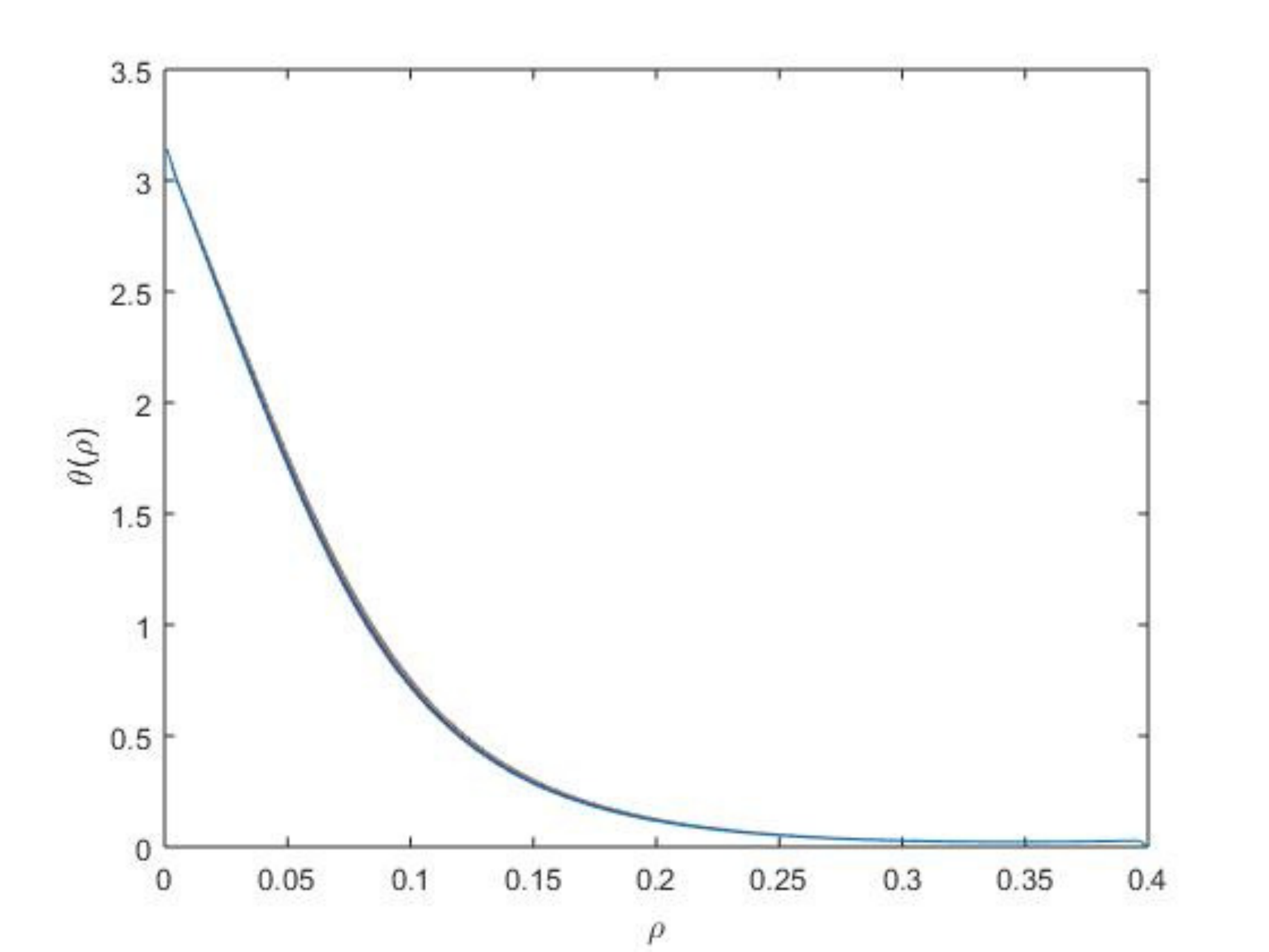}}
\subfigure[$\left(\dfrac{E}{E_0}\right) =1.5, k_s=6$]
{\includegraphics[width=.48\linewidth, height=5cm]{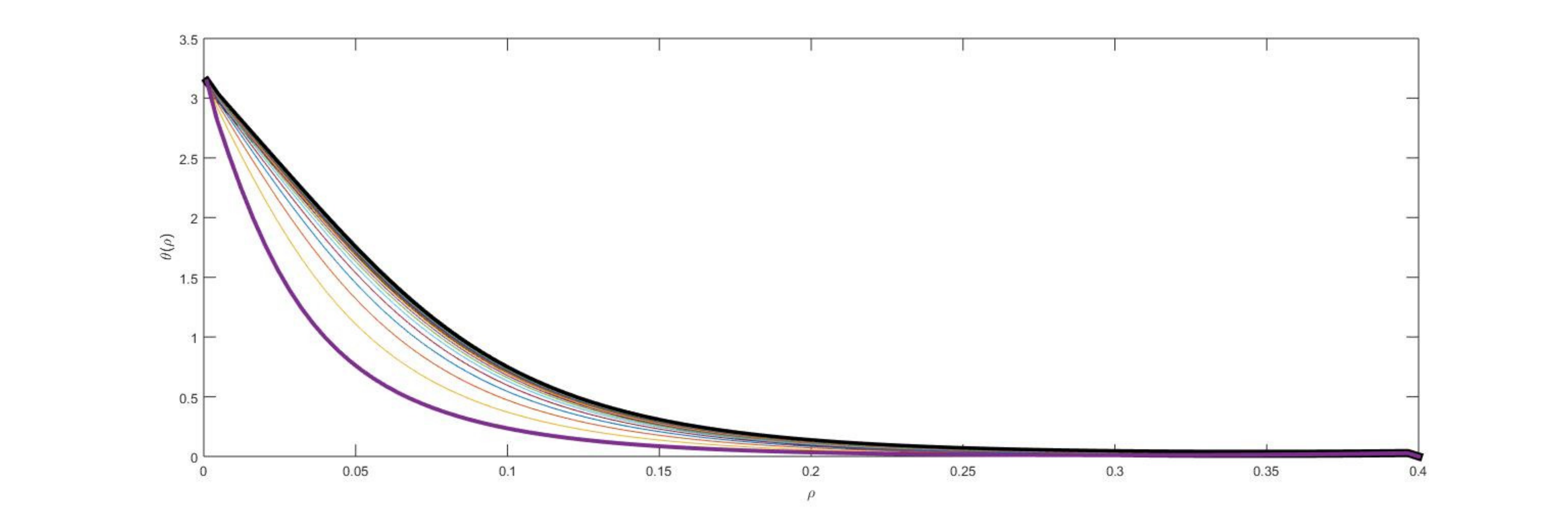}}
\caption{Profiles $\theta (\rho)$ for $E/E_0=1.5$. Different curves refer to different values of $\mid z\mid$.  Bold curves have to be referred  to $z =0$ (the black one) and to 
$\| z \|  =\nu/2$ (the purple one). We note that the effect of a greater external electric field is to narrow the size of the vortices, for fixed values of $k_s$.}
\label{fig:profiles2}
\end{figure}

 The numerical solutions of the BVP \eqref{theta2Dscal}, for different values of the couple $\left( \dfrac{E}{E_0}, k_s\right)$ are depicted in figures \ref{fig:profiles1} and \ref{fig:profiles2}. In each figure the profiles $\theta(\rho)$ for different values of $z\in [-\nu/2,\nu/2]$ are represented. In figure \ref{fig:profiles1} we have $\dfrac{E}{E_0}=1.02$ and the strength of the anchoring  $k_s=0.1, 6$. In figure \ref{fig:profiles2} we have $\dfrac{E}{E_0}=1.5$ with the same values of $k_s$. We note that, when the strength of the anchoring is small, the profiles are almost the same for every value of coordinate $z$. This means that, when the interfaces at the boundaries of the cell have a really small homeotropic effect on the director's configuration, a quasi-perfect cylindrical simmetry holds for axisymmetric solutions. In this case, the planar vortices described by $\theta(\rho)$ for every value of $z$, have the same, maximum, size. However, if we impose a quite stronger homeotropic effect at the boundaries, the vortices tend to have a reduced size, which becomes smaller as $\mid z\mid$ reaches the value $\dfrac{\nu}{2}$. In both figures \ref{fig:profiles1} and \ref{fig:profiles2}, the value of the adimensional thickness of the cell is $\nu=1.8$.
 
\begin{figure}
\centering
\subfigure[$\left(\dfrac{E}{E_0}\right) =1.02$]{\includegraphics[width=.6\linewidth,height=5.2cm]{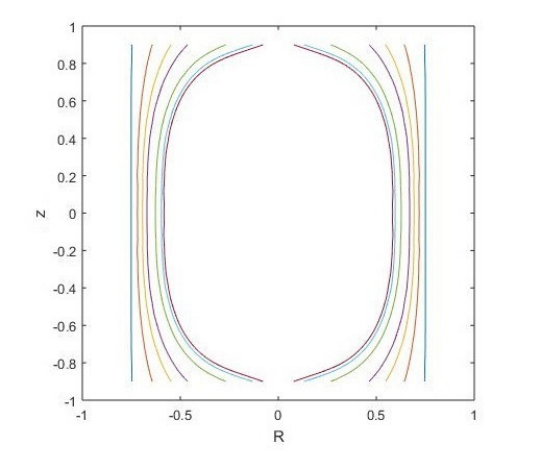}}
\subfigure[$\left(\dfrac{E}{E_0}\right) =1.5$]{\includegraphics[width=.6\linewidth,height=4.8cm]{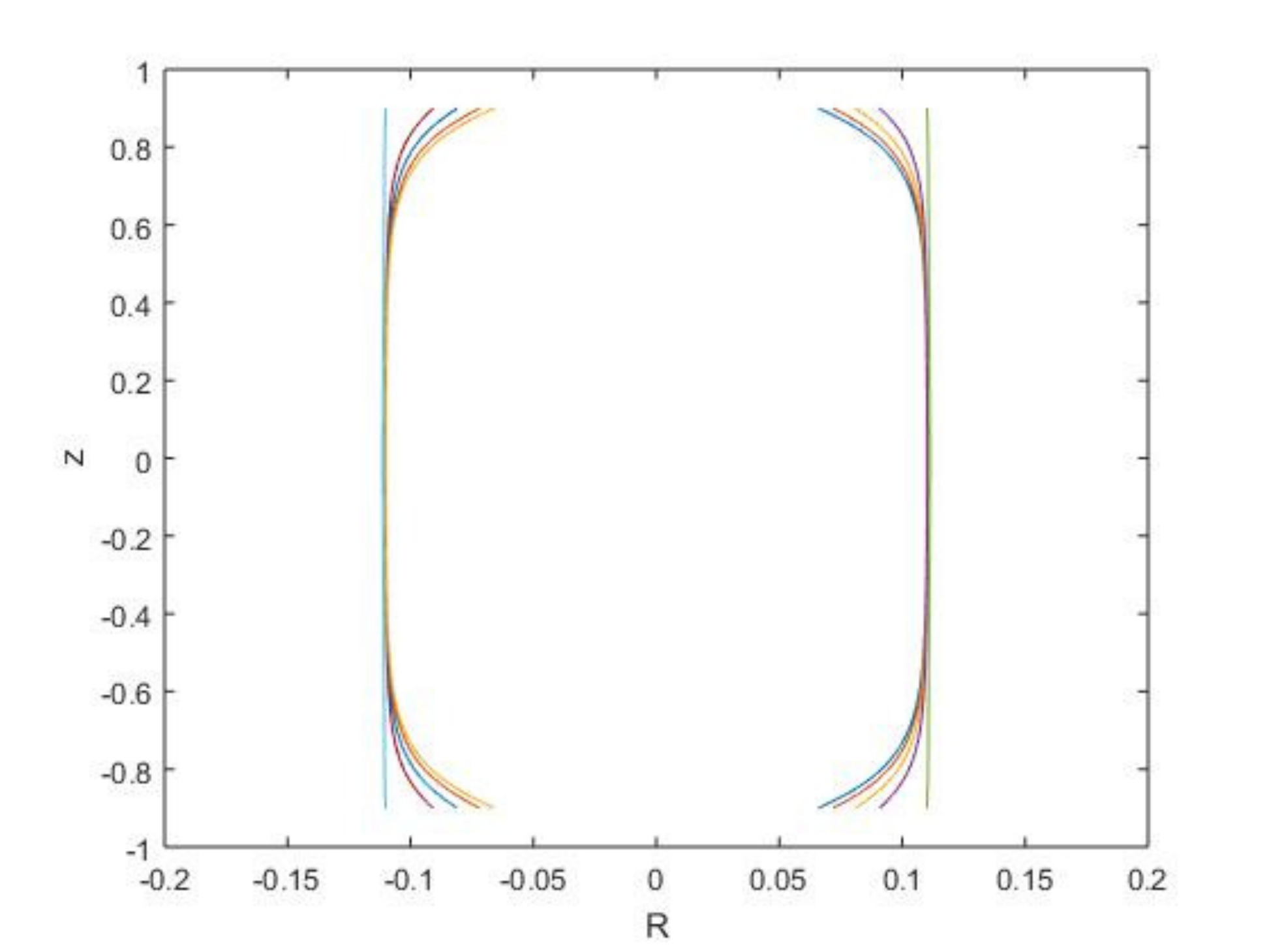}}
\caption{Size of the planar vortices for different values of $\mid z \mid $ . Different colors refer to different values of $k_s$ ($k_s=0.1,0.5,1,1.5,3,6,12$). We note that for smaller $k_s$ more similar to a cylinder are the solutions. The structures tend to take the form of a bubble as $k_s$ increases.}
\label{fig:size1}
\end{figure}

 A different representation of the spherulite is given by  reporting the intersection point with the $\rho$ axis by  the tangent to the inflection point of $\theta(\rho)$,  for a fixed value of $z$\cite{Leonov}. The results of this procedure are reported in figures \ref{fig:size1}.a and \ref{fig:size1}.b, for the two different values of $\dfrac{E}{E_0}$ taken into consideration. We stress that for greater external fields the size of all vortices narrows.

\section{Diffusion of Light on a CLC cylindrical structure }\label{sec:diffusion}
In this section we consider the scattering of an e.m. wave, propagating through  a confined CLC, which is under the suitable conditions for a spherulite to be formed. The geometry is the same as in the previous section, as the same is the choice of the cartesian axes. We assume that  the wave vector is parallel to the  $\lf  x, y\rg$ - plane. The propagation of the wave is described in terms of the oscillating electric field $\vE$, eventually to be distinguished by the static electric field $\mathbf{E}$, and by the associated electric displacement  field  $\vD$, by the equation \cite{Jackson}
\beq \nabla  \lf \nabla\cdot \vE \rg - \nabla^2 \, \vE = - \p_{t t } \vD. \label{Waveeq}\eeq
This equation has been obtained, as usual,  by eliminating the magnetic field from the Maxwell's equations. The electric anisotropy of the CLC is made explicit by the existence of a permeability tensor, which locally has an orthogonal component   $\epsilon_\bot$ if $\vE \bot \mathbf{n}$ and by a parallel contribution  $\epsilon_\|$ if $\vE \| \mathbf{n}$. Then the costitutive relation is given by \cite{deGennes,Kleman,Stewart}
\beq \vD = \epsilon_\bot \, \vE + \Delta \epsilon\; \mathbf{n}  \lf \vE \cdot \mathbf{n} \rg , \qquad  \Delta \epsilon =  \epsilon_\| - \epsilon_\bot .\eeq

Let us assume that  the incident wave is described by the electric field
\beq \vE \leadsto   E_y \; e^{\imath \lf k x -\omega t \rg}\bj + e^{\imath \lf \tilde{k} x -\omega t \rg}\bk  = {\vec \cE_\infty} e^{- \imath \omega t }\qquad x \freccia - \infty,\eeq where $\tilde{k} = k \sqrt{1+ \frac{\Delta \epsilon}{ \epsilon_\bot}}$.

Since we suppose that the spherulite is not perturbed by the wave, we need to assume certain supplementary conditions: 
\begin{enumerate}
\item the liquid crystal molecules are not deformed/rotated by wave, which implies $\omega  \gg \frac{1}{\tau}$, being $\tau$ any "relaxation time" of the CLC.
\item The diffractive effects in the light scattering on the spherulite are not negligible, then we assume that its wavelength is $\lambda \lesssim \rho_0 $ ( or $k\lf \omega \rg \gtrsim \frac{1}{\rho_0}$), being $\rho_0$ the typical size of the spherulite defined in equation \eqref{rho0}. 
\item the horizontal bounding plates are considered as  homogeneous dielectric planes, thus restricting the electric field to be periodic along the $z$ axis.
\item A strong supplementary condition we introduce is $\nabla \cdot \vE = 0$, which may imply $\nabla \cdot \vD = \rho_{free} \neq 0$. This should be true in the core of the spherulites, where we expect significant variations of the fields. However, at this  stage of our analysis we prefer to adopt  such an assumption, because the equations become simpler. Then an a posteriori evaluation of the local free charge density will clarify how good is our hypothesis. 
\item A final remark concerns the functional dependency of the shape of the spherulite, which we assume to be simply $\theta = \theta\lf \rho \rg$. Thus, for sake of simplicity we neglect the modulation along the $z$ axis  described by \eqref{zansatz}.
\end{enumerate}
Under the conditions above  equation \eqref{Waveeq} leads the equation for ${\vec \cE}= {\vec \cE}\lf \vr\rg$ 
\beq \label{redWaveeq} \nabla^2\; {\vec \cE} = - k^2 \cA\; {\vec \cE}, \quad  {\vec \cE}\leadsto  {\vec \cE_\infty} \; \textrm{as} \; x \to - \infty, \eeq
where $k = \frac{\omega}{c} \sqrt{\epsilon_\bot}$ and   the coupling matrix
\beq  \cA =  {\bf1}_3 + \frac{\Delta \epsilon}{\epsilon_\bot} \bn \otimes \bn. \eeq
Since  $\bn = \bn\lf \rho, \phi\rg =  \cos\theta\lf\rho\rg\bk  + \sin\theta\lf\rho\rg \boldsymbol{\phi} $, we are naturally led  to express also the electric field as 
\beq \vcE =  \cE_\rho \lf \rho, \phi, z \rg\,\boldsymbol{\rho}+  \cE_\phi  \lf \rho, \phi, z \rg\,\boldsymbol{\phi} + \cE_z  \lf \rho, \phi, z \rg\,\bk .\eeq
 Then, now it is easier to explicit  the off diagonal contributions to the \eqref{redWaveeq}. Indeed, provided that 
 \beq  \lf \bn \cdot \vcE \rg \, \bn=  \lf \sin \theta \,\cE_\phi + \cos \theta \,\cE_z\rg   \lf  \cos\theta\, \bk + \sin\theta\, \bphi  \rg, \eeq
 then \eqref{redWaveeq}  reads 
 \bea &\lf \nabla^2 + k^2 \rg \lf \cE_\rho \,\boldsymbol{\rho}+  \cE_\phi \,\boldsymbol{\phi} + \cE_z  \,\bk\rg = \label{vectorhelmh}\\ \nn
& - k^2 \frac{\Delta \epsilon}{\epsilon_\bot} \lq \lf \sin \theta \,\cE_\phi + \cos \theta \,\cE_z\rg  \cos\theta\, \bk + \lf \sin \theta \,\cE_\phi + \cos \theta \,\cE_z\rg \sin\theta\, \bphi \rq . \eea However, now the Laplacian operator acts on the cylindrical components of a vector-field, then it takes different expressions according to the component index. In particular, by defining $\nabla_0^2  \cdot = \frac{1}{\rho} \p_\rho \lf \rho \, \p_\rho \cdot  \rg + \frac{1}{\rho^2} \p^2_\phi + \p^2_z $, equation \eqref{vectorhelmh} becomes 
\bea \lf \nabla_0^2 + k^2 \rg   \cE_\rho & =&  \frac{1}{\rho^2} \lf  \cE_\rho  +2\, \p_\phi \, \cE_\phi  \rg  ,  \label{WaveRho}\\
 \lf \nabla_0^2 + k^2 \rg   \cE_\phi & =&  \frac{1}{\rho^2} \lf  \cE_\phi  - 2\, \p_\phi \, \cE_\rho  \rg - k^2 \frac{\Delta \epsilon}{\epsilon_\bot} \lf \sin^2 \theta \,\cE_\phi + \frac{1}{2} \sin 2\theta \,\cE_z\rg ,  \label{WavePhi}\\
 \lf \nabla_0^2 + \tilde{k}^2 \rg   \cE_z & = & - k^2 \frac{\Delta \epsilon}{\epsilon_\bot} \lf \frac{1}{2} \sin 2\theta \,\cE_\phi - \sin^2 \theta \,\cE_z\rg , \qquad \tilde{k} = k \sqrt{1 + \frac{\Delta \epsilon}{\epsilon_\bot}}. \label{WaveZ} \eea  In order to describe the scattering of the light on the spherulite, the above equations have to be solved with the asymptotic conditions 
 \beq  \cE_{ \rho} \leadsto \cE_{\infty \rho} \sin \phi \,e^{\imath {k} \rho \cos \phi }, \; \cE_{ \phi} \leadsto \cE_{\infty \phi} \cos \phi \,e^{\imath {k} \rho \cos \phi },\; \cE_{ z} \leadsto \cE_{\infty z} e^{\imath \tilde{k} \rho \cos \phi } \quad\textrm{for}\; \phi\to \pm\pi \; \textrm{and} \; \rho \to \infty.\label{asymptFields}\eeq
 Of course such asymptotic conditions  are exact solutions of the homogeneous system above, i.e. when $\frac{\Delta \epsilon}{\epsilon_\bot} \to 0$. As the problem of finding a complete analytical solution to \eqref{WavePhi} -  \eqref{WaveZ} is quite hard, let us consider a  perturbative setting. The basic idea is to  first give a Born approximated solution of the equation \eqref{WaveZ}, keeping an implicit dependence on $\cE_{ \phi}$. Then we can use it in \eqref{WavePhi} which will become a closed  linear equation, even if non local, in $\cE_{ \phi}$. Solving it, in the same approximation, one can use these results into \eqref{WaveRho}  for $\cE_{ \rho}$. 
 \section{Perturbative solutions of the light scattering equations by a spherulite}\label{sec:Born}
 \subsection{The \emph{out plane} conversion} \label{sec:Bornout}
Following the standard method by Lippmann-Schwinger \cite{LippSchwi},  let us rewrite  equation \eqref{WaveZ} as the integral equation 
 \beq  \cE_z\lf \vr\rg = \cE_{\infty z} e^{\imath \tilde{k} \rho \cos \phi }  + \int G\lf \vr,\vr'\rg U\lq \cE_z\lf \vr'\rg, \cE_\phi\lf \vr'\rg,\theta\lf \rho'\rg \rq \; d\vr' \label{LSintEq}\eeq
 where $U\lq \cE_z\lf \vr\rg, \cE_\phi\lf \vr\rg,\theta\lf \rho\rg \rq  = - k^2 \frac{\Delta \epsilon}{\epsilon_\bot} \lf \frac{1}{2} \sin 2\theta\lf \rho\rg \,\cE_\phi \lf \vr\rg - \sin^2 \theta\lf \rho\rg \,\cE_z\lf \vr\rg \rg  $ and the Green function $G\lf \vr,\vr'\rg$ is a solution of the PDE 
 \beq   \lf \nabla_0^2 + \tilde{k}^2 \rg G\lf \vr,\vr'\rg = \frac{1}{\rho'} \delta\lf \rho-\rho'\rg \delta\lf \phi-\phi'\rg \delta\lf z - z'\rg  \label{Geq}\eeq provided that it is  differentiable in its  domain (i.e. the CLC layer) except at the point $\rho = \rho', \; \phi = \phi' , \; z = z'$. There the partial first derivatives exist, but they are not continuous,  in such a way that the second derivatives admit the singularity defined by the r.h.s. in \eqref{Geq}.

 The function $G$ can take the form
 \beq G = \frac{1}{2 \pi \nu} \sum_{m, \, n = -\infty}^\infty e^{\frac{2 \pi \imath n}{\nu} \lf z - z'\rg} e^{ \imath m \lf \phi - \phi'\rg}\, h_{m, \, n }\lf \rho, \rho' \rg ,\label{GreenSeries}\eeq
 where the functions $h_{n,m}$ satisfy the Bessel type equation with singular inhomogeneity \cite{NIST:DLMF}
 \beq \lq \frac{1}{\rho}\, \p_\rho \lf \rho \, \p_\rho\, \cdot\rg - \frac{m^2}{\rho^2} + \tilde{k}^2 - \lf\frac{2 \pi n}{\nu}\rg^2 \rq \, h_{m, \, n} = \frac{1}{\rho'} \delta\lf \rho - \rho'\rg,\label{BesselEq}\eeq and the cutoff frequency $\kappa_n = \sqrt{\tilde{k}^2 - \lf\frac{2 \pi n}{\nu}\rg^2} \; \textrm{for} \; \tilde{k} \geq \frac{2 \pi n}{\nu}.$ is induced by the finite transverse size of the CLC layer.  

 We require $G$ to be a continuous function  with a bounded behaviour at  $\rho \to 0$ and, additionally,  to be a cylindrical progressive wave as  $\rho \to \infty$, i.e.  of the form $\propto \frac{e^{\imath \kappa \rho}}{\sqrt{\kappa \rho}}$. Furthemore,  because of the $\delta$ like inhomogeneity, $G$ can have discontinuities only in the first derivatives at $\rho \to \rho'$.

In conclusion, by imposing the above conditions,  the Green function \eqref{GreenSeries} takes the form
\beq G\lf \vr,\vr'\rg  =  \frac{- \imath}{4  \nu} \sum_{m, \, n = -\infty}^\infty e^{\frac{2 \pi \imath n}{\nu} \lf z - z'\rg} e^{ \imath m \lf \phi - \phi'\rg}\, H^{\lf 1 \rg}_m\lf \kappa_n \, \rho_> \rg J_m\lf \kappa_n \, \rho_< \rg,\qquad 
\left\{
\begin{array}{ccc}
\rho_>  &   = & \textrm{max}\lf \rho, \rho'\rg  \\
    \rho_< &  = &  \textrm{min}\lf \rho, \rho'\rg 
\end{array}
\right. \eeq
where $J_m$ denotes the Bessel function of first kind with integer order $m$ and $H^{\lf 1 \rg}_m \lf \zeta \rg = J_m\lf \zeta \rg + \imath\, Y_m\lf \zeta \rg$ the corresponding Hankel function of first kind \cite{NIST:DLMF}.  Without further calculations,   dramatic simplifications stem from our assumption 5. in Sec. 3, implying that the only non vanishing contributions come from the $n = 0$ mode. 
 Moreover, we are actually interested in the behaviour of the wave at  radii much larger than the effective size of the spherulite, which decreases very fast, as we noticed in (2.14). Thus the form of Green function we have to use is 
 \beq G_{sempl}\lf \vr,\vr'\rg  =  \frac{- \imath}{4} \sum_{m  = -\infty}^\infty e^{ \imath m \lf \phi - \phi'\rg}\, H^{\lf 1 \rg}_m\lf \tilde{k} \, \rho \rg J_m\lf \tilde{k} \, \rho'\rg, \eeq

Now,  replacing the above formula into \eqref{LSintEq} and introducing the explicit form of the potential $U$, we see from \eqref{WaveZ}  that  the parameter  $\frac{\Delta \epsilon}{\epsilon_\bot}$ can be considered as a perturbation parameter, allowing to  express the wave function as a series of powers of it.  At the 0 order the solution is given by  asymptotics \eqref{asymptFields}, which replaced into \eqref{LSintEq} provides at  the first order (Born approximation) corrections  to the plane wave propagation. 
 
  Thus, in the Born limit, by the identity $ e^{\imath \zeta \cos p} = \sum_{l = -\infty}^\infty e^{\frac{\imath \,\pi \,l}{2}} \,e^{\imath \, l \, p}\, J_l\lf\zeta\rg$, one can integrate on $\phi'$ and obtain   the  approximated expression $ \cE_z^B\lf \vr\rg $ of the  $ \cE_z\lf \vr\rg $ component  as follows
  

    \bea & \cE_z^{B}\lf \vr\rg = \cE_{\infty z} e^{\imath \tilde{k} \rho \cos \phi }
  +  \frac{\pi}{2}\,\frac{\Delta \epsilon  k^2 }{ 
   \epsilon _\bot} \sum_{m = -\infty}^\infty e^{i m \phi}
 H_m^{(1)}( \tilde{k}\,\rho) \nn \\ &\int\lq \cE_{\infty \phi
   }\frac{\imath^{m}}{4} \,\sin (2
   \theta\lf \rho'\rg)J_m( \tilde{k}\, \rho')\lf J_{m-1}( k\, \rho') - J_{m+1}( k\, \rho') \rg  - \cE_{\text{$\infty $z}} \imath^{m+1}\,\sin
   ^2(\theta \lf \rho'\rg)J_m^2( \tilde{k}\, \rho')  \rq \; \rho' d\rho'  .    \label{BornZaxis2}   \eea
   
   In order to have a simple estimation of the integrals in the above expression, let us resort to the asymptotic expressions of the spherulite given by \eqref{ansatzbulk0} and \eqref{asymptlin}. Actually, the simplest rough choice is \eqref{linansatz} (with $Z\lf z \rg = 1$), which we will adopt here, since we are not interested in the exact values of the diffusion amplitudes, but only in their approximate size. Thus, we have to evaluate integrals of the form 
   \bea  {\cal I}^\phi_m&=&  - \int_0^{\pi  k \rho_0 }\, \sin \lf 2
  \frac{ s}{ k \rho_0 } \rg J_m( s)\lf J_{m-1}( s) - J_{m+1}( s) \rg \; s\; d s \, = \nn \\
  & \,&  - \int_0^{\pi  k \rho_0 }\, \sin \lf 2
  \frac{ s}{ k \rho_0 } \rg \lf J_m(s)^2 \rg'  \; s\; d s, \label{matrix1}\\
 {\cal I}^z_m&=&   \int_0^{\pi k \rho_0}\, \sin
   ^2\left(  \frac{ s}{ k \rho_0 } \right)J_m^2(  s)  \; s \;d s,  \label{matrix2}\eea
   where the substitution $\tilde{k} \to k$ is justified,  since the difference is of the order $ \frac{\Delta \epsilon  }{ 
   \epsilon _\bot}$ as stated in \eqref{WaveZ}.
   
   At the moment the above matrix elements  do  not have yet  an analytical expression and should be computed numerically.    

\begin{figure}
\centering
{\includegraphics[width=.6\linewidth,height=5.2cm]{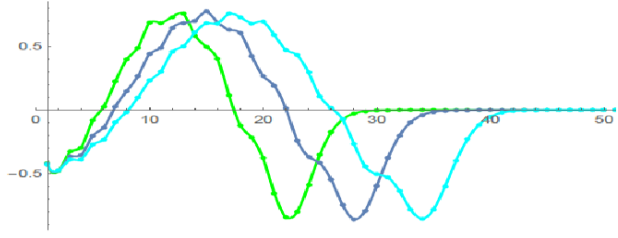}}
\caption{The numerical values of ${\cal I}^\phi_m$ as function of $0 \leq m \leq  50$ for three different values of $k \rho_0$, precisely 8 (green), 10 (blue) and 12 (cyan). They can be approximated by continuous functions in $m$, for instance by linear  combinations of gaussian functions, but still a clear pattern for a systematic approximation has to be developed. It is clear a linear dependence, with a  coefficient $\sim 4$, of the number of significantly terms with the size parameter $k \rho_0$. This is in agreement with the scattering features on localized central potentials.   }
\label{fig:integrIm}
\end{figure}

Examples of the numerical evaluation of  a certain number of integrals \eqref{matrix1} is given in figure \ref{fig:integrIm}. 
   \bigskip
   
   Before  proceeding  in such a calculations let us show the form of the cross section of conversion of a \emph{in plane} polarized wave into a \emph{out plane } polarized one. In fact let us  suppose $\cE_{\infty z} = 0$, then the scattered amplitude along the $z$-axis \eqref{BornZaxis2}  reads
   \bea
    \cE_z^{B}\lf \vr\rg =
  \, \cE_{\infty \phi} \frac{\pi\, \Delta \epsilon   }{ 
  8 \, \epsilon _\bot}  \sum_{m = -\infty}^\infty  \imath^{m} \, {\cal I}^\phi_m \, e^{i m \phi}
 H_m^{(1)}( \tilde{k}\,\rho) . \nn
   \eea
   Recalling that at infinity the asymptotic behaviour of the Hankel functions is $ H_m^{(1)}( \zeta) \leadsto \frac{(1-i) e^{i \zeta-\frac{i \pi
    m}{2}}}{\sqrt{\pi\, \zeta }} + O\left(\zeta^{-3/2}\right),$ the above expression becomes 
    \bea
    \cE_z^{B}\lf \vr\rg &=&
  \lf 1 - \imath\rg \, \cE_{\infty \phi} \frac{\pi\,\Delta \epsilon   }{ 
  8\, \epsilon _\bot} \frac{e^{ \imath\, \tilde{k}\,\rho }}{\sqrt{\pi \,  \tilde{k}\,\rho}} \sum_{m = -\infty}^\infty   \, {\cal I}^\phi_m \, e^{i m \phi}
   =  \nn \\
 &  &\lf 1 - \imath\rg \, \cE_{\infty \phi} \frac{\pi\,\Delta \epsilon   }{ 
  8\, \epsilon _\bot} \frac{e^{ \imath\, \tilde{k}\,\rho }}{\sqrt{\pi \,  \tilde{k}\,\rho}}\, \lq {\cal I}^\phi_0 + 2 \sum_{m = 1 }^{+ \infty} \, {\cal I}^\phi_m \, \cos{ m \phi} \rq ,
   \eea
   where the identity ${\cal I}^\phi_{- m} = {\cal I}^\phi_m$ has been used, which can be easily proved from  \eqref{matrix1} and by $J_{- m} = \lf -1\rg^m \, J_m$.

The cross section of the conversion of  linear \textit{in plane}  polarized $\hy$ light into the \textit{out plane} $\hz$ one is given by 
\beq 
\frac{d\, \sigma_{conv}}{d \phi}\lf \hr, \hz; \hx, \hy \rg =  \frac{\pi}{32}\,\sqrt{\frac{\epsilon_\bot}{\epsilon_{||}}} \lf \frac{\Delta \epsilon   }{ 
  \epsilon _\bot} \rg^2 \,  \frac{\nu\, \rho_0}{ k\, \rho_0} \,
\lq {\cal I}^\phi_0 + 2 \sum_{m = 1 }^{+ \infty} \, {\cal I}^\phi_m \, \cos{ m \phi} \rq^2,
\label{convcrsection}\eeq
where we have singled out the dependency on the geometrical size of the spherulites from its  relative size with respect the used light wavelength.

The calculations of the conversion cross section in the direction $\hr\lf \phi \rg$ indicates that there is a quite well defined small angle,  around $10^o$ in our numerical examples, along which the rotation of the polarization is efficiently performed. The  angle of maximum conversion is   $\propto \lf k \rho_0\rg ^{-1}$, thus it becomes  smaller as   the wavelength becomes shorter.  A  further remarkable aspect is the vanishing of the backscattering. The effective values depend basically on the square of the anisotropy ratio  
$\lf \frac{\Delta \epsilon   }{ 
   \epsilon _\bot} \rg^2$. In fact the total cross section takes the expression 
  \beq 
 \sigma_{conv} = \frac{\pi^2}{16}\,\sqrt{\frac{\epsilon_\bot}{\epsilon_{||}}} \, \lf \frac{\Delta \epsilon   }{ 
   \epsilon _\bot} \rg^2\,\frac{\nu\, \rho_0}{ k \rho_0}\,
\lq  \lf {\cal I}^\phi_0\rg^2 + 2 \sum_{m = 1 }^{ \infty} \,\lf {\cal I}^\phi_m\rg^2
\rq , \eeq
which is a decreasing function of $k\, \rho_0$.   

\begin{figure}
\centering
{\includegraphics[width=.65\linewidth,height=8.3cm]{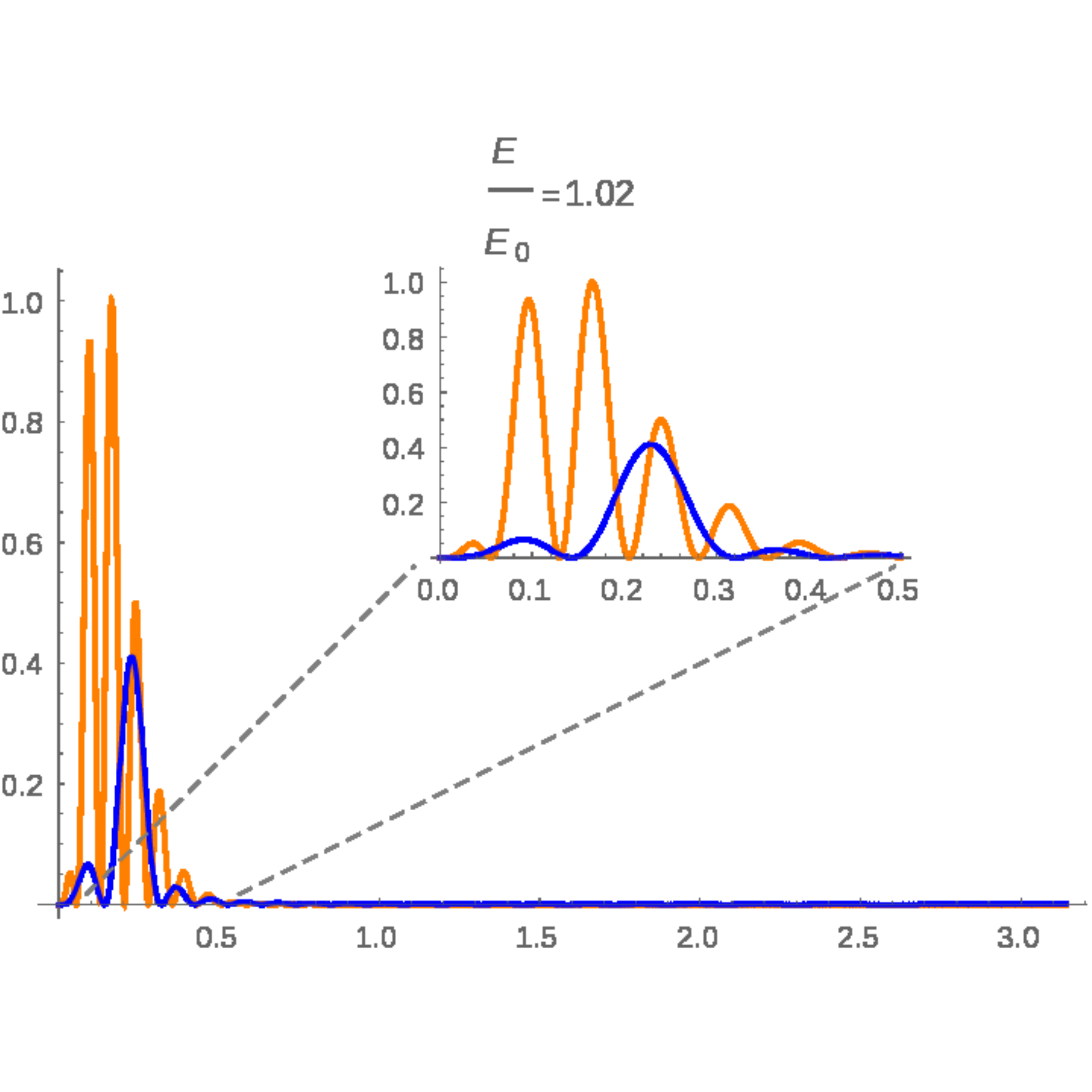}}
\caption{The numerical evaluation of the conversion cross section \eqref{convcrsection}, in arbitrary units, for $\rho_0$ fixed by \eqref{rho0} and the ratio $E/E_0 = 1.02$,  $k \rho_0 = 10$ and for the exact numerical solution (orange) and the approximated solution \eqref{ansatzbulk0} (blue).}
\label{fig:comparison1}
\end{figure}

\begin{figure}
\centering
{\includegraphics[width=.65\linewidth,height=8.3cm]{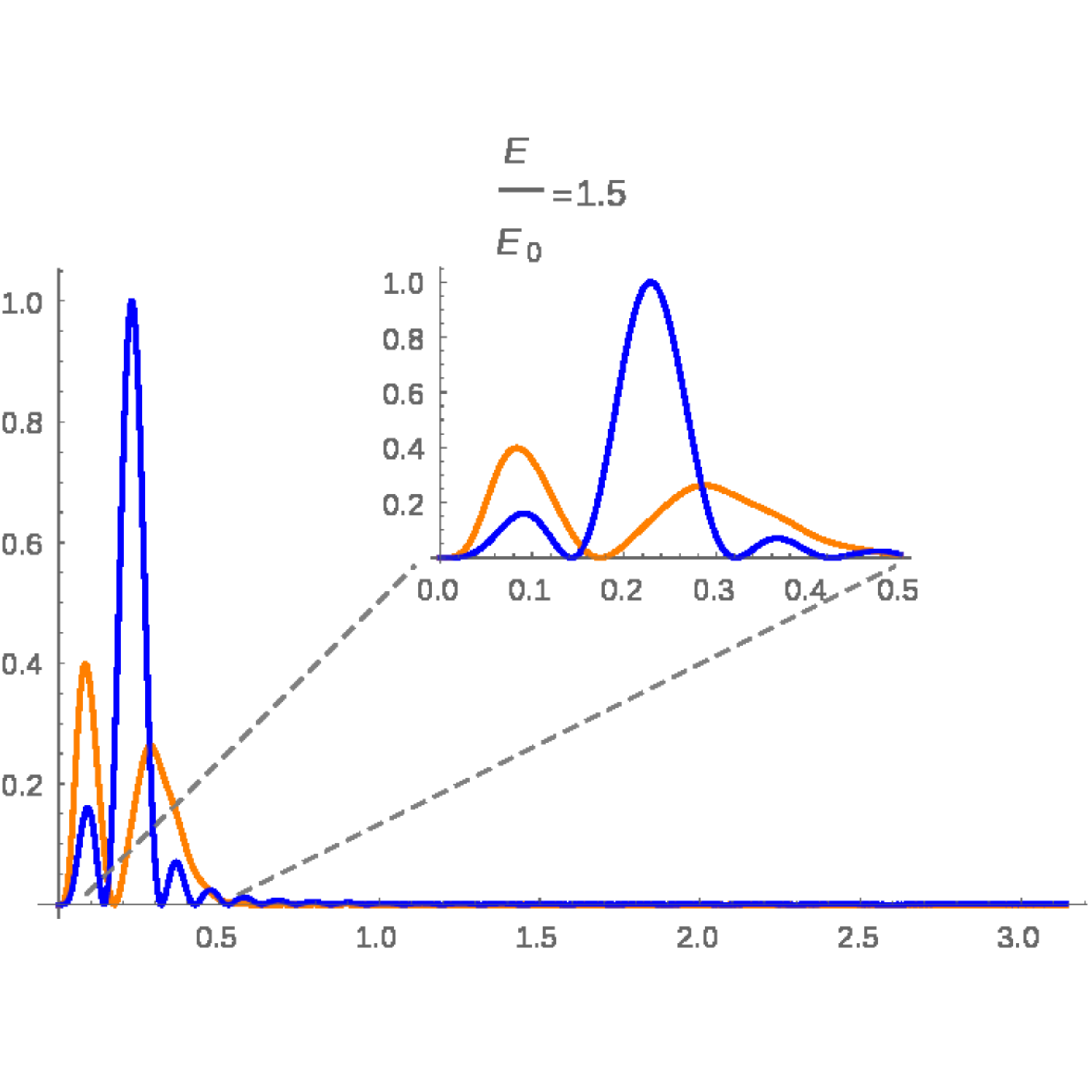}}
\caption{The numerical evaluation of the conversion cross section \eqref{convcrsection}, in arbitrary units, for $\rho_0$ fixed by \eqref{rho0} and the ratio $E/E_0 = 1.5$,  $k \rho_0 = 10$ and for the exact numerical solution (orange) and the approximated solution \eqref{ansatzbulk0} (blue).}
\label{fig:comparison2}
\end{figure}

Let us turn our attention again to equation \eqref{BornZaxis2}. By using the exact numerical solutions for the spherulite (figures \ref{fig:profiles1} and \ref{fig:profiles2}), we can obtain the exact expression of  the differential cross section in \eqref{convcrsection}, for the scattering of an electromagnetic wave by a skyrmion.  A direct comparison between the exact differential cross section and the approximated one in arbitrary units is reported in  figures \ref{fig:comparison1} and \ref{fig:comparison2}. 
As it can be seen, the exact numerical solution  for the spherulites makes the angle of maximum conversion smaller,  than the one computed through the use of the approximated solution.  Furthermore, recalling that as the external electric field increases, the size of the spherulite decreases (as described by equation \eqref{rho0}), we note that the larger is the size of the skyrmion, more  efficient  is the polarization conversion with respect to the approximated one.
    \begin{figure}[h]
\centering
{\includegraphics[width=.5\linewidth,height=6.2cm]{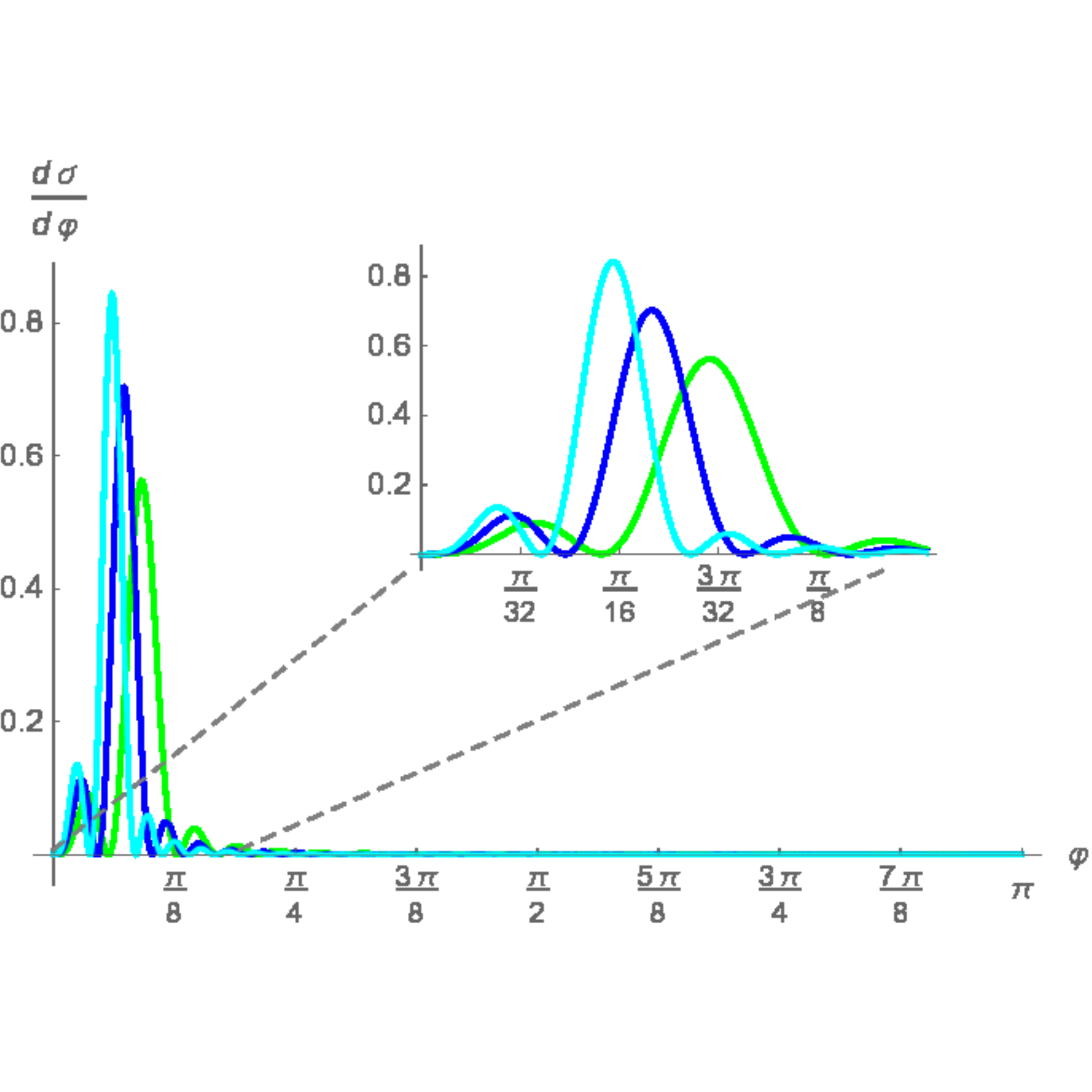}}
\caption{The numerical evaluation of the conversion cross section \eqref{convcrsection}, in arbitrary units,  for the three different values of $\tilde{k} \rho_0 = 8, 10, 12$, by using the same color code as in fig. \eqref{fig:integrIm}.}
\label{fig:bornconversion}
\end{figure}
 \subsection{The \emph{in plane}  conversion}\label{sec:Bornin}
 Now let us turn our attention on the subsystem \eqref{WaveRho}-\eqref{WavePhi}, which could be represented in the form
 \beq 
\left(
\begin{array}{cc}
 L + k^2 & -M      \\
 M  & L   + k^2   
\end{array}
\right) \left(
\begin{array}{c}
 \cE _\rho     \\
 \cE _\phi     
\end{array}
\right) =    k^2 \frac{\Delta \epsilon}{\epsilon_\bot} \lf  \begin{array}{c}
 0     \\
  \sin^2 \theta \,\cE_\phi + \frac{1}{2} \sin 2\theta \,\cE_z     
\end{array}
\rg  , \label{Erhophisub}
\eeq
where $L =   \nabla_0^2  - \frac{1}{\rho^2} $ and $M =   \frac{2}{\rho^2}\, \p_\phi \,   $ .
As before,  it can be set into the integral form
\bea 
&\lf \begin{array}{c}
 \cE _\rho     \\
 \cE _\phi     
\end{array}
\right)  &  =  \lf \begin{array}{c}
  \cE_{\infty \rho} \sin \phi \\
 \cE_{\infty \phi} \cos \phi 
\end{array}
\right) \,e^{\imath {k} \rho \cos \phi }  +  \nn \\
  &k^2 \frac{\Delta \epsilon}{\epsilon_\bot} & \int \left(
\begin{array}{cc}
 G\lf \vr,\vr'\rg & - F\lf \vr,\vr'\rg      \\
 F\lf \vr,\vr'\rg  & G\lf \vr,\vr'\rg      
\end{array}
\right) \lf  \begin{array}{c}
 0     \\
  \sin^2 \theta\lf \vr'\rg \,\cE_\phi \lf \vr'\rg+ \frac{1}{2} \sin 2\theta\lf \vr'\rg \,\cE_z\lf \vr'\rg     
\end{array}
\rg   \; d\vr' .
\eea
Of course, the inhomogeneous term is a solution of the homogeneous system \eqref{Erhophisub}  and the matrix Green function 
 is 
\beq  \left(
\begin{array}{cc}
 G\lf \vr,\vr'\rg & - F\lf \vr,\vr'\rg      \\
 F\lf \vr,\vr'\rg  & G\lf \vr,\vr'\rg      
\end{array}
\right) =  \frac{1}{2 \pi \nu} \sum_{m, \, n = -\infty}^\infty e^{\frac{2 \pi \imath n}{\nu} \lf z - z'\rg} e^{ \imath m \lf \phi - \phi'\rg}\,   \left(
\begin{array}{cc}
 h_{m, \, n }\lf \rho, \rho' \rg & - \imath\,f_{m, \, n }\lf \rho, \rho' \rg    \\
\imath\, f_{m, \, n }\lf \rho, \rho' \rg  & h_{m, \, n }\lf \rho, \rho' \rg      
\end{array}
\right) ,  \eeq
where the unknown  $ h_{m, \, n },  \,f_{m, \, n }$ satisfy the matrix equation
\beq \label{systemhf}
\lf \begin{array}{cc}
\frac{1}{\rho}\, \p_\rho \lf \rho \, \p_\rho\, \cdot\rg - \frac{m^2+1}{\rho^2} +\kappa_n^2& \frac{-2 \imath \, m}{\rho^2}    \\
\frac{2 \imath \, m}{\rho^2} & \frac{1}{\rho}\, \p_\rho \lf \rho \, \p_\rho\, \cdot\rg - \frac{m^2+1}{\rho^2} +\kappa_n^2      
\end{array}
\right) \left(
\begin{array}{cc}
 h_{m, \, n }\lf \rho, \rho' \rg & -\imath\, f_{m, \, n }\lf \rho, \rho' \rg    \\
\imath\, f_{m, \, n }\lf \rho, \rho' \rg  & h_{m, \, n }\lf \rho, \rho' \rg      
\end{array}
\right) =   \,\frac{\mathbf{ 1}_2 }{\rho'} \delta\lf \rho - \rho'\rg,
\eeq
where $\kappa_n^2= k^2 -\lf \frac{2\, \pi \, n}{\nu}\rg^2$.

As in the previous subsection, we  limit ourselves  to evaluate the diffusion of light by the spherulite in the Born approximation. Accordingly, the conversion from out-plane to in-plane scattering leads to the following approximated  expression 
\beq 
\lf \begin{array}{c}
 \cE _\rho^B     \\
 \cE _\phi ^B    
\end{array}
\right)  = \cE_{\infty z}\, \frac{k^2 }{2 \pi \nu} \frac{\Delta \epsilon}{2\,\epsilon_\bot}  \sum_{m, \, n = -\infty}^\infty \int  \sin 2\theta\lf \rho'\rg \, e^{\imath \tilde{k} \rho' \cos \phi'}\, e^{\frac{2 \pi \imath n}{\nu} \lf z - z'\rg} 
e^{ \imath m \lf \phi - \phi'\rg}\,   
\left(
\begin{array}{c}
  -\imath\, f_{m, \, n }\lf \rho, \rho' \rg    \\
 h_{m, \, n }\lf \rho, \rho' \rg      
\end{array}
\right) \, d\vr'   ,
\eeq
where $h_{m, \, n }$ and $f_{m, \, n }$ are solutions of the system \eqref{systemhf}. 
Again, using the simplification induced by the assumption 5.  in Sec. 3 and by using the expansion of the plane wave factor in terms of Bessel functions,
\; one gets 
\beq 
\lf \begin{array}{c}
 \cE _\rho^B     \\
 \cE _\phi ^B    
\end{array}
\right)  = \cE_{\infty z}\,  \frac{\Delta \epsilon\,k^2 }{2\,\epsilon_\bot}  \sum_{m\,= -\infty}^\infty \imath^m\,e^{ \imath m \, \phi} \int  \sin 2\theta\lf \rho'\rg  \, J_m\lf  \tilde{k} \rho' \rg \, 
\,   
\left(
\begin{array}{c}
  -\imath\, f_{m }\lf \rho, \rho' \rg    \\
 h_{m }\lf \rho, \rho' \rg      
\end{array}
\right) \, \rho'\,  d\rho'  , \label{InplaneBorn}
\eeq
where we dropped the subscript $n$ from both $h_{m, \, n }$ and $f_{m, \, n }$ as the only non vanishing contributions come from the $n = 0 $ mode. 
 The squared  modulus of the above quantity, properly managed, will produce the cross section of the  \emph{out plane} - \emph{in plane }  scattering process.
 
 From \eqref{systemhf}  we obtain   the equations for $h_{m}$ and $f_{m}$  as follows
 \beq \left\{\begin{array}{cc} h_m''\lf \rho \rg+\frac{h_m'\lf \rho \rg}{\rho}+ \left(k^2-\frac{m^2+1}{\rho^2} \right) h_m\lf \rho \rg +\frac{2 mf_m\lf \rho \rg}{\rho^2}  & = \frac{\delta\lf \rho - \rho' \rg}{\rho'} \, ,   \\f_m''\lf \rho \rg+\frac{f_m'\lf \rho \rg}{\rho}+\left(k^2-\frac{m^2+1}{\rho^2}
   \right) f_m\lf \rho \rg +\frac{2 m h_m\lf \rho \rg}{\rho^2}   & = 0.
   \end{array}\right. \label{syshf}
 \eeq

   The general solution of the system above is
   \bea  h_m^\pm \, &=& c_1^\pm \, J_{m-1}\lf k\, \rho \rg + \imath c_2^\pm \, Y_{m-1}\lf k\, \rho \rg +  d_1^\pm \, J_{m+1}\lf k\, \rho \rg+ \imath  d_2^\pm \,Y_{m+1}\lf k\, \rho \rg,\nn \\
   f_m^\pm \, &= & c_1^\pm \, J_{m-1}\lf k\, \rho \rg + \imath  c_2^\pm \, Y_{m-1}\lf k\, \rho \rg -  d_1^\pm \, J_{m+1}\lf k\, \rho \rg - \imath d_2^\pm \, Y_{m+1}\lf k\, \rho \rg,
   \eea
   where $c_i^\pm \,$ and $d_i^\pm \,$ are four arbitrary constants in the regions $\rho <\rho' $ or $\rho > \rho' $, respectively. Continuity of the solutions  and discontinuity of their first derivatives at $\rho'$  imply   a functional dependency of those coefficients on this variable.  Moreover, as in the previous section,  we require regularity at $\rho \to \, 0$ and radiative behaviour at $\rho \to \, \infty$.

  All  conditions above lead to a  linear system, from which  one obtains the values of the unknown coefficients, namely
   \bea c_1^{+} = -\frac{\imath\, \pi}{4} H^{\lf 1 \rg}_{m-1}\lf k\, \rho' \rg, \;  
   d_1^{+} =  - \frac{\imath\, \pi}{4} H^{\lf 1 \rg}_{m+1}\lf k\, \rho' \rg, \;  c_2^{+} = d_2^{+} = 0 \; , \nn \\
c_1^{-} = c_2^{-} = -\frac{\imath\, \pi}{4} J_{m-1}\lf k\, \rho' \rg, \;  d_1^{-} =  d_2^{-} =  - \frac{\imath\, \pi}{4} J_{m+1}\lf k\, \rho' \rg . \eea
  Now we are in position to evaluate \eqref{InplaneBorn}, namely
  \small{
  \beq 
  \frac{ \cE_{\infty z}\,  \pi\, \Delta \epsilon\,k^2 }{8\,\epsilon_\bot}  \sum_{m\,= -\infty}^\infty \imath^{m-1}\,e^{ \imath m \, \phi} \int  \sin 2\theta\lf \rho'\rg  \, J_m\lf  \tilde{k} \rho' \rg \, 
\,   
\left(
\begin{array}{c}
  -\imath\, \lq J_{m-1}\lf k\, \rho' \rg H^{\lf 1 \rg}_{m-1}\lf k\, \rho \rg -  J_{m+1}\lf k\, \rho' \rg  H^{\lf 1 \rg}_{m+1}\lf k\, \rho \rg \rq    \\
 J_{m-1}\lf k\, \rho' \rg H^{\lf 1 \rg}_{m-1}\lf k\, \rho \rg +  J_{m+1}\lf k\, \rho' \rg  H^{\lf 1 \rg}_{m+1}\lf k\, \rho \rg      
\end{array}
\right) \, \rho'\,  d\rho'  . \label{InplaneBorn1}
\eeq}
Resorting again to the asymptotic behaviour of the Hankel functions, the solution at infinity can be estimated as
 \beq 
  \frac{ \cE_{\infty z}\, \sqrt{ \pi}\, \Delta \epsilon\,k^2 }{ 2^{\frac{5}{2}}\,\epsilon_\bot}   e^{-\imath\, \frac{\pi}{4}}\frac{e^{\imath \, k \,\rho}}{\sqrt{k \,\rho}}\sum_{m\,= -\infty}^\infty \,e^{ \imath m \, \phi} \int  \sin 2\theta\lf \rho'\rg  \, J_m\lf  \tilde{k} \rho' \rg \, 
\,   
\left(
\begin{array}{c}
  -\imath\, \lq J_{m-1}\lf k\, \rho' \rg  +  J_{m+1}\lf k\, \rho' \rg   \rq    \\
 J_{m-1}\lf k\, \rho' \rg  -  J_{m+1}\lf k\, \rho' \rg       
\end{array}
\right) \, \rho'\,  d\rho'   . \label{InplaneBorn2}
\eeq

Setting

\begin{eqnarray}
 \frac{1}{\tilde{k}^2} I^{(\rho)}_m=&\frac{1}{k^2} \int  \sin 2\theta\lf \frac{s}{\tilde{k}}\rg  \, J_m\lf  s \rg \, 
\,   
 \lq J_{m-1}\lf k\, \frac{s}{\tilde{k}} \rg  +  J_{m+1}\lf k\, \frac{s}{\tilde{k}} \rg   \rq      
 \, s\,  d s\, d\phi' \label{matrix3}\\
 \frac{1}{\tilde{k}^2}I^{(\phi)}_m=&\frac{1}{k^2}\int  \sin 2\theta\lf \frac{s}{\tilde{k}}\rg  \, J_m\lf  s \rg \, 
\,   
 \lq J_{m-1}\lf k\,\frac{s}{\tilde{k}} \rg  -  J_{m+1}\lf k\, \frac{s}{\tilde{k}} \rg   \rq      
 \, s\,  d s,
 \label{matrix4}
\end{eqnarray}

equation \eqref{InplaneBorn2} can be rewritten as
\beq
  \frac{ \cE_{\infty z}\, \sqrt{ \pi}\, \Delta \epsilon }{ 2^{\frac{5}{2}}\,\epsilon_\bot}   e^{-\imath\, \frac{\pi}{4}}\frac{e^{\imath \, k \,\rho}}{\sqrt{k \,\rho}}\sum_{m\,= -\infty}^\infty \,e^{ \imath m \, \phi} \left(\begin{array}{c}
  -\imath \,I^{(\rho)}_m\\ I^{(\phi)}_m
\end{array}
\right)   . \label{InplaneBorn3}
\eeq

\begin{figure}
\centering
\subfigure[$\dfrac{E}{E_0}=1.02$]{\includegraphics[width=.45\linewidth,height=5.3cm]{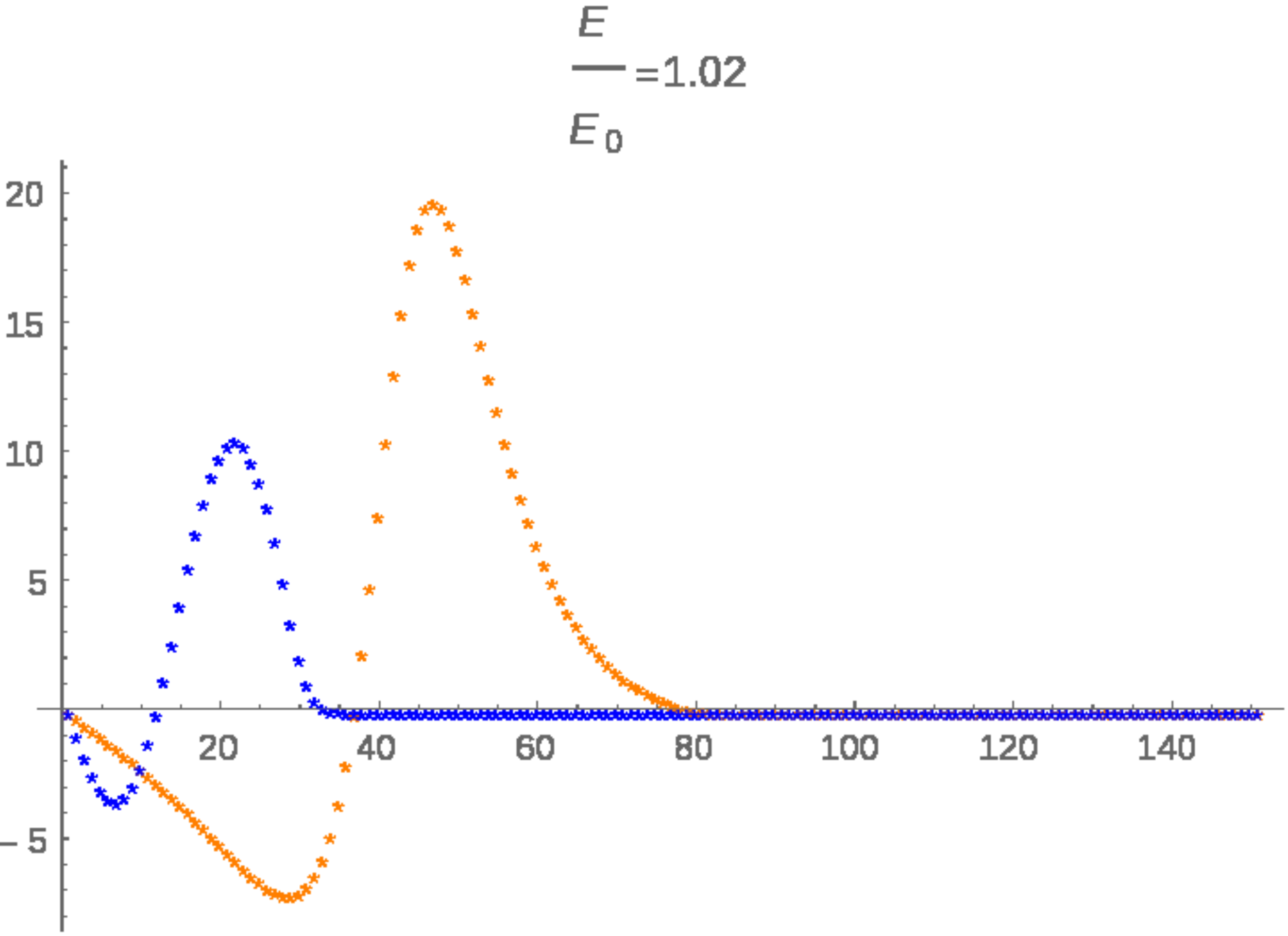}}
\subfigure[$\dfrac{E}{E_0}=1.5$]{\includegraphics[width=.45\linewidth,height=5.3cm]{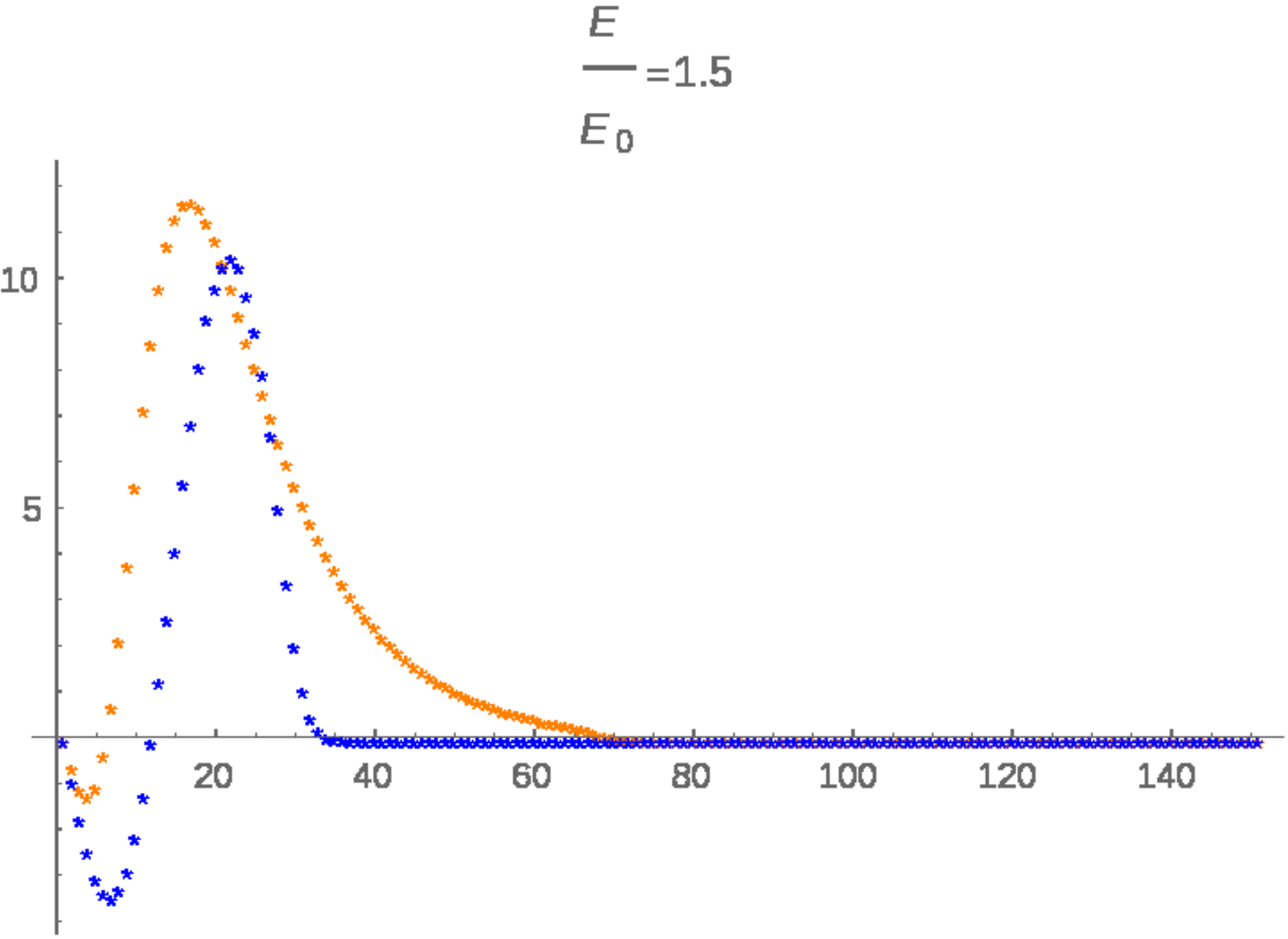}}
\caption{The numerical values of $I^{(\rho)}_m$ as function of $0 \leq m \leq  150$, for two different values of $E/E_0$ with fixed $k\rho_0$, computed through the use of the exact numerical solution (orange) and the approximated solution \eqref{ansatzbulk0} (blue).}
\label{fig:matrix3}
\end{figure}

Recalling the identity $J_{-m}=(-1)^m J_{m}$, it is easy to show that $I^{(\rho)}_0=0$, $I^{(\rho)}_m=-I^{(\rho)}_{-m}$  and $I^{(\phi)}_m=I^{(\phi)}_{-m}$, so that equation \eqref{InplaneBorn3} now reads 
\beq
  \frac{ \cE_{\infty z}\, \sqrt{ \pi}\, \Delta \epsilon }{ 2^{\frac{5}{2}}\,\epsilon_\bot}   e^{-\imath\, \frac{\pi}{4}}\frac{e^{\imath \, k \,\rho}}{\sqrt{k \,\rho}}
\left(
\begin{array}{c}
2 \sum_{m\,= 1}^\infty \, I_m^{(\rho)}\sin m\phi   \\
I^{(\phi)}_0+2 \sum_{m\,= 1}^\infty \, I_m^{(\phi)}\cos m\phi   
\end{array}
\right)  . \label{InplaneBorn4}
\eeq
Performing again the substitution $\tilde{k} \to k$, we notice that $I_{m}^{(\phi)}$ is the same as $\mathcal{I}_m^\phi$ in equation \eqref{matrix1}. On the other hand, the values of the first three hundred matrix elements \eqref{matrix3} are presented in figure \ref{fig:matrix3}.

 The \textit{in plane-conversion} cross section is then given by 
\beq 
\frac{d\, \sigma}{d \phi}\lf \hr, \hat{\phi}; \hx, \hy \rg =  \frac{\pi}{32}\,\sqrt{\frac{\epsilon_\bot}{\epsilon_{||}}} \lf \frac{\Delta \epsilon   }{ 
  \epsilon _\bot} \rg^2 \,  \frac{\nu\, \rho_0}{ k\, \rho_0} \,
\lq 4\left(\sum_{m\,= 1}^\infty \, I_m^{(\rho)}\sin m\phi\right)^2+\left({ I}^{(\phi)}_0 + 2 \sum_{m = 1 }^{+ \infty} \, { I}^{(\phi)}_m \, \cos{ m \phi}\right)^2 \rq
\label{convcrsection2}\eeq
and the total cross section reads
\beq 
\sigma_{\text{tot}}\lf \hr, \hat{\phi}; \hx, \hy \rg =  \frac{\pi^2}{16}\,\sqrt{\frac{\epsilon_\bot}{\epsilon_{||}}} \lf \frac{\Delta \epsilon   }{ 
  \epsilon _\bot} \rg^2 \,  \frac{\nu\, \rho_0}{ k\, \rho_0} \,
\lq\left({ I}^{(\phi)}_0\right)^2+ 2\sum_{m\,= 1}^\infty \, \left(I_m^{(\rho)}\right)^2 + 2 \sum_{m = 1 }^{+ \infty} \, \left({ I}^{(\phi)}_m \, \right)^2 \rq.
\label{convcrsectiontot2}\eeq

\begin{figure}
\centering
{\includegraphics[width=.65\linewidth,height=8.3cm]{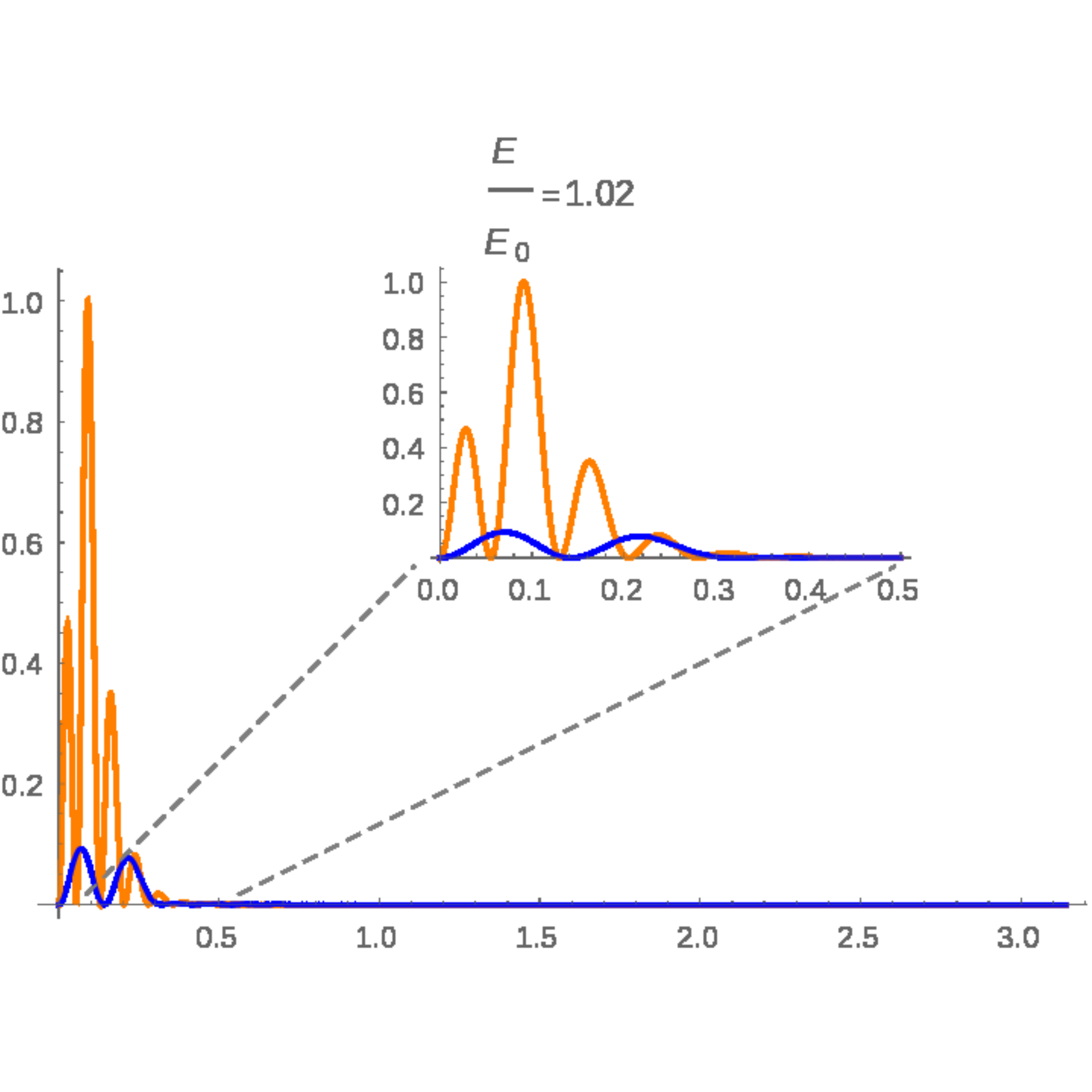}}
\caption{The numerical evaluation of the conversion cross section \eqref{convcrsection2}, in arbitrary units, for $\rho_0$ fixed by \eqref{rho0} and the ratio $E/E_0 = 1.02$,  $k \rho_0 = 10$ and for the exact numerical solution (orange) and the approximated solution \eqref{ansatzbulk0} (blue).}
\label{fig:sigmaplane}
\end{figure}

\begin{figure}
\centering
{\includegraphics[width=.65\linewidth,height=8.3cm]{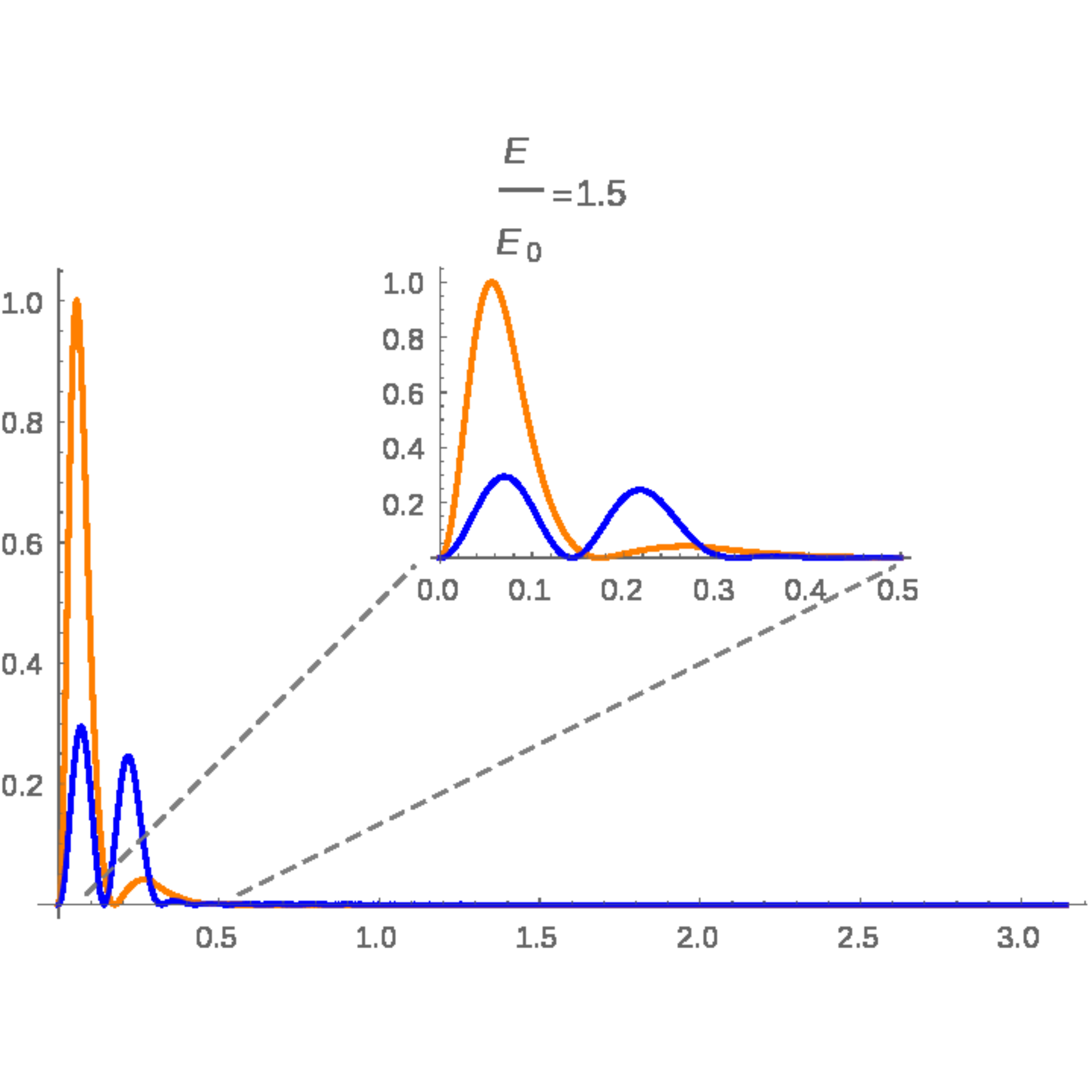}}
\caption{The numerical evaluation of the conversion cross section \eqref{convcrsection2}, in arbitrary units, for $\rho_0$ fixed by \eqref{rho0} and the ratio $E/E_0 = 1.5$,  $k \rho_0 = 10$ and for the exact numerical solution (orange) and the approximated solution \eqref{ansatzbulk0} (blue).}
\label{fig:sigmaplane2}
\end{figure}

The numerical results, in arbitrary units, for the computation of the differential cross section \eqref{convcrsection2} are depicted in figures \ref{fig:sigmaplane} and \ref{fig:sigmaplane2}, for two different values of the ratio $E/E_0$. Conversely to what happens for the \textit{out plane}-conversion cross section, in this case the use of the exact solution for the computation of the differential cross section \eqref{convcrsection2} keeps the angle of maximum conversion substantially unchanged.

\section{Conclusions}
In the present work we showed that the spherulites in CLC can be used to change the polarization axes of incoming light with a certain efficiency. To the best of our knowledge, this phenomenon is quite new  as, so far, only the light diffusion from helicoidal CLC structures in the bulk has been studied \cite{nadina}:  here we considered the interaction with localized perturbations, i.e. the spherulites.  In detail we first described the shape of the spherulites, for different values of the controlling parameters, in particular the external applied electric (or magnetic) field. From that we were able to compute the cross section of the polarization axes conversions in Born approximation. We found  that the conversion  processes have maximum differential cross section at small non-zero deflection  angles. Thus, the effect we described can be detected off  the  forward direction. 
Furthermore, we compared the differential cross sections  for different values of the external electric field, proving that the scattering in significantly influenced by such a parameter. Thus, we can use it as a tuning controller of the diffusion. In particular,  the conversion is more efficient for fields  slightly above the threshold of the critical unwinding field of the cholesteric-nematic transition. This is due to the quadratic inverse dependency on the external field of the spherulite core size. In order to obtain these results, we used both a piecewise linear approximation of the spherulite profile and the corresponding numerical exact solution.    On the other hand, we showed that the spherulite is badly approximated by a piecewise linear function, especially  for weak electric field. Thus, to improve our results we need to further study analytical profile solution of the spherulites . Actually there are many similar questions to be answered. First, it would be important to study the cross sections for all channels, beyond the Born approximation, and to suppress the several  simplifications we made. In particular, the spherulite is not a cylinder, as we assumed in the present work, but it resembles more a sort of barrel. Correspondingly,  new diffractive effects may arise from the actual shape, especially close  the confining plates. This is related to the type of anchoring, which is parametrised by a further controlling parameter in the Rapini-Papoular conditions. In fact,  we showed that the shape of the spherulite depends significantly on it, even if the  ratio $E/E_0$ is kept fixed.   Finally, it is well known that for  external fields  below the critical threshold, lattice configurations of spherulites can appear \cite{Leonov}. This fact  suggests to explore the light diffusion processes in such a regime, in order to enhance the effects we described above, or to have a better control on them.  

\subsubsection*{Aknowledgments}
This work was partially supported by MIUR,  by the INFN on the project IS-CSN4 Mathematical Methods of Nonlinear Physics and by INDAM-GNFM.

\begin{figure}
\center
{\includegraphics[width=\linewidth,height=8.3cm]{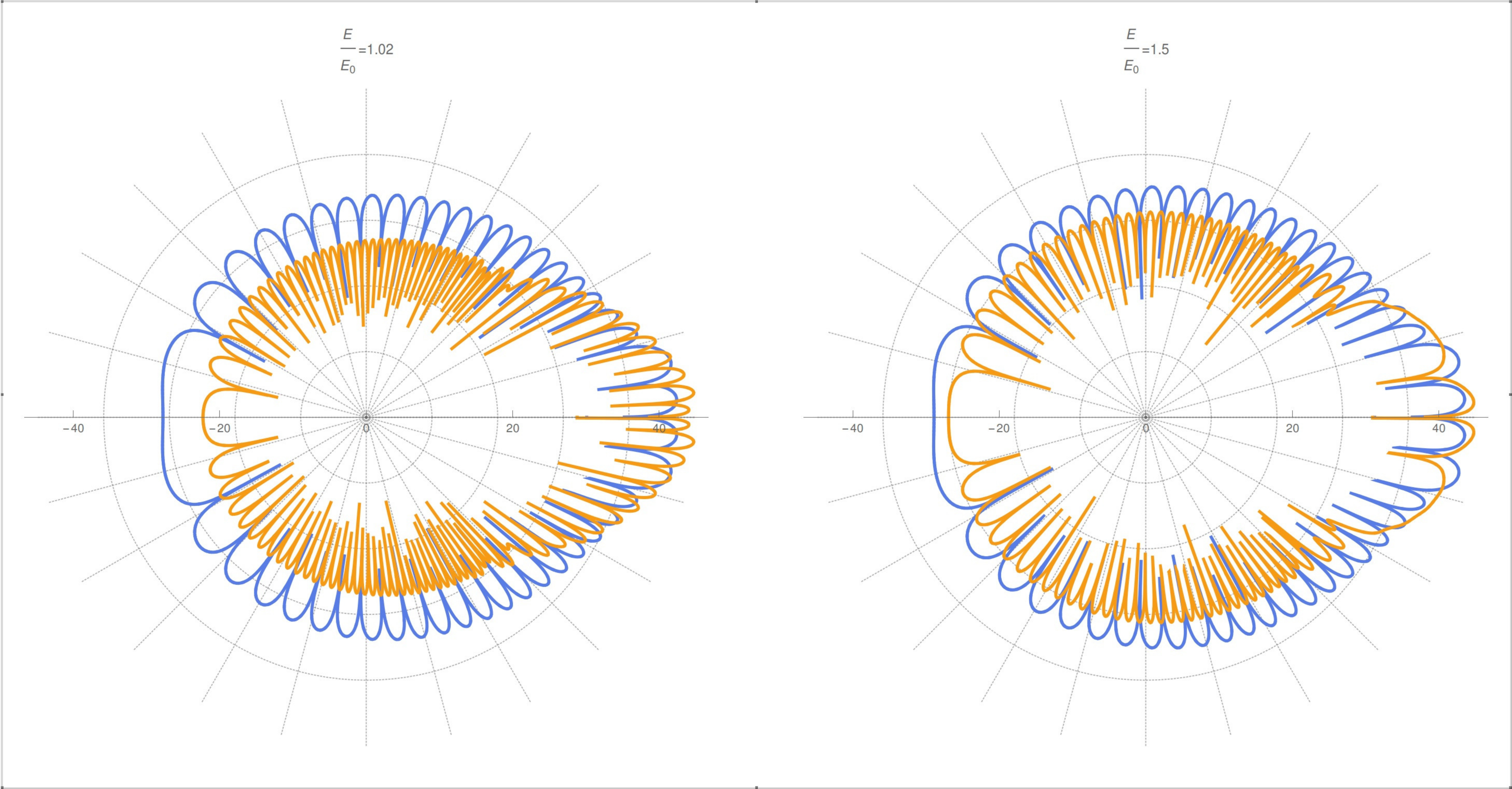}}
\caption{Comparison of the Log-polar plots of the differential cross sections for two different values of the external electric field.  In both cases
 the orange curves refer to  the exact numerical spherulite profile, while the  blue ones correspond to the piecewise linear approximation of it. }
\label{fig:polarsigma}
\end{figure}

\end{document}